\documentclass[prb,aps,twocolumn]{revtex4}
\usepackage{amsmath}
\usepackage[dvips]{graphicx}
\usepackage{subfigure}
\usepackage{epsfig}
\usepackage{graphics}

\newcommand{\be}{\begin{equation}}
\newcommand{\ee}{\end{equation}}
\newcommand{\bk}{{{\bf{k}}}}
\newcommand{\bK}{{{\bf{K}}}}

\newcommand{\bQ}{{{\bf{Q}}}}
\newcommand{\br}{{{\bf{r}}}}

\newcommand{\bq}{{\bf{q}}}

\newcommand{\bea}{\begin{eqnarray}}
\newcommand{\eea}{\end{eqnarray}}
\newcommand{\ra}{\rangle}
\newcommand{\la}{\langle}
\newcommand{\upa}{\uparrow}
\newcommand{\dna}{\downarrow}

\newcommand{\bT}{{\bf T}}

\newcommand{\dg}{{\dagger}}
\newcommand{\bPi}{{\boldsymbol \Pi}}
\newcommand{\pdg}{{\phantom\dagger}}
\newcommand{\nn}{{\nonumber}}

\begin{document}
\title{Collective modes and superflow instabilities of strongly correlated Fermi superfluids}
\author{R. Ganesh, A. Paramekanti}
\affiliation{Department of Physics, University of Toronto, Toronto,
Ontario M5S 1A7, Canada}
\author{A. A. Burkov}
\affiliation{Department of Physics, University of Waterloo, Waterloo,
Ontario N2L 3G1, Canada}
\begin{abstract}
We study the superfluid phase of the one-band attractive Hubbard model of fermions as a
prototype of a strongly correlated s-wave fermion superfluid on a lattice. We show that
the collective mode spectrum of this superfluid exhibits, in addition to the long wavelength
sound mode, a
sharp roton mode over a wide range of densities and interaction strengths. We compute
the sound velocity and the roton gap within a generalized random phase
approximation (GRPA) and show that the GRPA results are in good agreement, at strong coupling, 
with a spin wave analysis of the appropriate strong coupling
pseudospin model. We also investigate, using this two-pronged approach, the 
breakdown of
superfluidity in the presence of a supercurrent. We find that the superflow can break down
at a critical flow momentum
via several distinct mechanisms --- depairing, Landau instabilities or dynamical instabilities
--- depending on the dimension, the interaction
strength and the fermion density. The most interesting of these instabilities is a charge modulation 
dynamical instability which is
distinct from previously studied dynamical instabilities of Bose superfluids. The charge order
associated with this instability can be of two types: (i) a commensurate checkerboard modulation
driven
by softening of the roton mode at the Brillouin zone corner, or, 
(ii) an incommensurate density modulation arising from superflow-induced
finite momentum pairing of Bogoliubov quasiparticles. We elucidate
the dynamical phase diagram showing the critical flow momentum of the leading
instability over a wide
range of fermion densities and interaction strengths and point out
implications of our results for experiments on cold atom fermion superfluids in an optical lattice.
\end{abstract}
\maketitle
\section{Introduction}
The study of strongly correlated fermionic superfluids is of interest in the context of solid 
state materials, such as cuprate \cite{WenLee} and pnictide superconductors,\cite{norman08}
as well as ultracold
atomic gases in optical lattices.~\cite{Bloch08,lehur09} 
One of the key goals of condensed matter physics is to understand
the nature of single particle and collective excitations in such strongly correlated systems. 
Another important problem is to
eludicate the mechanisms by which superfluidity breaks down in these lattice systems ---
such an understanding would shed light on the critical current and 
on the nature of vortex cores in strongly interacting superfluids,~\cite{Aeppli01} 
which are issues
of significant experimental and theoretical interest. 
Moreover, the inverse critical superflow momentum can be shown to be directly related to 
characteristic length 
scales, associated with 
the low-energy excitations in the system and can thus serve as a useful
probe of these excitations. This is especially relevant for neutral cold atom superfluids, in which 
the critical flow is unaffected by complications such as disorder effects and current induced 
magnetic fields which are present in  solid state superconductors.
While there has been a significant
amount of experimental and theoretical progress in understanding the 
breakdown of
superflow for Bose superfluids in an optical lattice,~\cite{Niu01,Altman05,nikuni06,ketterle-bose07} and for Fermi superfluids in the absence of a lattice
potential,~\cite{sensarma06,Ketterle07} 
this issue has not been addressed in detail in the context of Fermi 
superfluids on a lattice.  

In this paper we study the collective modes and the mechanisms for the 
breakdown of superflow in the one-band attractive Hubbard model which is a model Hamiltonian 
for strongly correlated s-wave Fermi superfluids on a lattice.~\cite{Ranninger90,swz,
trivedi,levin,Scalapino91}
The attractive Hubbard Hamiltonian,
\bea
\label{eq:1}
H &=& -t \sum_{\la ij \ra} (c_{i\sigma}^\dagger c^{\vphantom \dag}_{j\sigma} +
c_{j\sigma}^\dagger c^{\vphantom \dag}_{i\sigma}) - \mu\sum_{i\sigma}n_{i\sigma} \nonumber \\
&-& U\sum_{i}(n_{i\uparrow}-\frac{1}{2}) (n_{i\downarrow}-\frac{1}{2}),
\label{eq:hub}
\eea
describes fermions with spin $\sigma$ hopping, with amplitude $t$, between adjacent sites of a 
lattice, and interacting via an on-site attractive
interaction $U$. The chemical potential $\mu$ tunes the fermion density away from half-filling at $\mu=0$.
Summation over repeated spin indices is implicit henceforth, and, unless
stated explicitly, we will measure energies in units of $t$.
We will study this model on the square lattice in 
two dimensions (2D) and the cubic lattice
in three dimensions (3D). The ground states of this model include the uniform
superfluid (SF) and a ``checkerboard'' charge density wave (CDW), which is an
insulating crystal.~\cite{Ranninger90}
It is well known that this model has an enhanced ``pseudospin symmetry''  
at the point $\mu=0$,~\cite{Zhang90} which leads to a
degeneracy between the SF and the CDW ground states.
Away from $\mu=0$, this degeneracy is lifted and the SF ground 
state is energetically more favorable.
One  expects, however, that so long as this pseudospin symmetry is only
weakly broken, the excitation spectrum of the
superfluid will exhibit signatures of proximity to the
CDW ground state. One also
expects that once superflow is imposed, there will be
a critical flow velocity beyond which the kinetic energy of superflow will 
overcome the energy difference between the stationary SF and the CDW.
The superflow in this case is limited not by the pairing gap, which may be large, but by this 
energy difference, which becomes very small close to half-filling. 
This situation is similar to the one encountered, for example, in cuprate superconductors, 
where  the
superconducting ground state in the underdoped region of the phase diagram is proximate 
to the antiferromagnetic Mott insulator.~\cite{MacDonald08}  Similar physics may also be relevant to
Cu$_x$TiSe$_2$ where there is a competition between charge order and superconductivity.~\cite{morosan06}

This general idea motivates us to study the collective modes as well
the stability of superflow in the SF phase of the attractive Hubbard model.
We begin by studying the collective mode spectrum in the absence of superflow
and find that there is indeed a large window of interactions
and density where the collective mode spectrum exhibits, in addition
to a sound mode, a well-defined roton
minimum at wavevectors $\bPi \equiv (\pi,\pi)$ (in 2D) or $\bPi
\equiv (\pi,\pi,\pi)$ (in 3D) arising
from proximity to the CDW. 
The existence of such a roton
minimum has been pointed out in earlier work.\cite{kostyrko,alm,danshita}
We present our results for the values of sound velocity and roton gap as functions of
fermion filling and interaction strength. 
These results of the collective
mode spectrum could potentially be verified in studies of collective modes
in the superfluid phase of cold Fermi gases in an optical lattice.
We then turn to a study of the stability of uniform superflow in the 
Hubbard model.

We elucidate a variety of superflow-breakdown mechanisms - depairing, Landau instabilities
and dynamical instabilities. We observe Landau instabilities of the collective mode at incommensurate
wavevectors as has also been 
reported in Ref.[\onlinecite{danshita}]. More importantly, we discover
dynamical instabilities involving checkerboard or incommensurate stripe-like density
modulations which are distinct from previously studied dynamical instabilities of
Bose superfluids.~\cite{Niu01,Altman05} We show that the checkerboard dynamical instability can be 
viewed as the result of the softening of the roton mode and that this instability
has a simple analog in the strong coupling pseudospin model,~\cite{Burkov08}
which we discuss. 
In contrast, the incommensurate CDW instability occurs only at intermediate coupling 
strengths
and has no analog in the strong-coupling pseudospin model. We explain this instability 
as a finite momentum pairing instability 
of Bogoliubov quasiparticles, somewhat analogous to the Halperin-Rice exciton
condensate instability in
indirect band-gap semiconductors.\cite{Halperin1968}
These
dynamical instabilities could be experimentally probed by creating a
``running'' optical lattice as has been done for Bose 
superfluids.~\cite{ketterle-bose07}

The existence of
the checkerboard dynamical instability was discussed earlier by us from the strong coupling
perspective.~\cite{Burkov08}  The present paper goes considerably beyond our earlier work.
The formalism that we use (the generalized random
phase approximation or GRPA), allows us to address the
entire range of interaction strengths from weak to strong coupling. We have checked that
the GRPA results smoothly match on to the earlier analysis of the strong coupling
pseudospin model and we present these comparisons where appropriate. Within this
formalism we have obtained dynamical phase diagrams of the flowing superfluid over
a wider range of interactions than in our previous study and uncovered the incommensurate
CDW instability which was not reported in our earlier paper.

We have tried to make this paper fairly self-contained.
We begin
in Section II with an overview of some key general results on the Hubbard model.
Section III focuses on the formalism which we use in the remainder of the paper.
We then turn, in Section IV, to results for the collective mode spectrum in the absence
of superflow. Section V contains the results for the instabilities of the flowing
superfluid and the phase
diagram showing the leading instabilities as a function of interaction and fermion
density. Section VI contains a discussion of the experimental observability of our results in 
fermionic cold atom systems as well as possible implications of these results for vortex core physics
in strongly correlated superfluids. We end, in Section VII, with a summary of our results and
possible avenues for future research. Appendix A contains the Fourier transform convention which
we use, Appendix B gives details of the GRPA formalism, and the 
mean field theory of the ``flowing supersolid'' state is discussed in Appendix C.

\section{Overview of the Model}
Before we turn to a study of collective modes and the stability of superflow using
the Hubbard model Hamiltonian, we begin by reviewing some well known facts
about the model
which will serve to set the stage for our discussion in the remainder of this paper.

\subsection{Pseudospin Operators and Pseudospin Symmetry}

We start by defining pseudospin operators, first introduced in the context of 
the theory of superconductivity by Anderson:~\cite{Anderson58}
\bea
\label{eq:2}
T_i^+ &=& \eta_i c_{i\uparrow}^\dagger c_{i\downarrow}^\dagger, \nonumber \\
T_i^- &=& \eta_i c_{i\downarrow}c_{i\uparrow}, \nonumber \\
T_i^z&=&\frac{1}{2}(c_{i\sigma}^\dagger c^{\vphantom \dag}_{i\sigma}-1),
\eea
where $\eta_i=+ 1$ on
one sublattice and $\eta_i=-1$ on the other sublattice of the square or cubic lattice.
The physical meaning of these operators is clear: $T^{+}_i$ creates a
fermion pair at site $i$, $T^-_i$ annihilates a fermion pair at site $i$,
and $T^z_i$ is the pair number operator. 
It is straightforward to check that these operators obey usual spin commutation 
relations. Furthermore, provided that $\mu=0$,
the global pseudospin operators,
\bea
\label{eq:3}
T^z &=& \sum_i T^z_i, \nonumber \\
T^{\pm} &=& \sum_i  T^{\pm}_i, 
\eea
all commute with the Hubbard Hamiltonian in Eq.~(\ref{eq:hub}) so that there is a global
pseudospin SU(2) symmetry at this special point.~\cite{Zhang90} This is in addition to the ordinary global 
spin SU(2) symmetry which is present for any value of $\mu$.
Pseudospin language is often very convenient and intuitive, in particular when discussing symmetry properties 
and collective modes of paired-fermion superfluids.  

\subsection{Ground States and Degeneracies}
Quantum Monte Carlo simulations~\cite{Scalapino91,swz,trivedi} 
have shown that the ground state of the attractive 
Hubbard model is a uniform SF
for generic values of $U/t$ and $\mu/t$. The uniform SF has
an order parameter $\la c_{i\uparrow}^\dagger c_{i\downarrow}^\dagger\ra
\sim \Delta {\rm e}^{i\varphi}$ where a choice of the
phase, $\varphi$, of the order parameter corresponds to a spontaneously broken
symmetry. In terms of pseudospin operators, this means that
$\la T^+_i \ra \sim \eta_i \Delta {\rm e}^{i\varphi}$.
For $\mu=0$, the pseudospin symmetry discussed above leads to the conclusion that
all states related to this SF ground state by a global pseudospin rotation
are also valid ground states and are characterized by an order parameter 
${\bf N} = \eta_i \la {\bf T}_i \ra$ which can point to any location on the Bloch sphere.
The uniform SF state has ${\bf N}$ pointing to locations on the equator.
The state with ${\bf N}
\sim \pm \Delta \hat{z}$, which points to the north/south pole of the Bloch sphere,
corresponds to a checkerboard charge density wave state (CDW) of pairs. 
Other locations on the Bloch sphere correspond to states with coexisting CDW and SF orders.

\subsection{Strong coupling pseudospin Hamiltonian}
In the limit of large $U/t$, each site on the lattice
will have either no fermions or a pair of tightly
bound fermions. Thus, any allowed fermion configuration will have $T^z_i = \pm 1/2$ 
at each site and the kinetic
energy will lead to tunneling between such ``classical'' Ising configurations of the
pseudospin. The pseudospin dynamics for $U/t \gg 1$ is then described by an
effective pseudospin Hamiltonian:
\be
\label{eq:4}
H_{\rm eff} = J  \sum_{<ij>}{\bf T}_i \cdot {\bf T}_j - \mu \sum_i T_i^z, 
\ee
where $J = 4 t^2 /  U$. 
For $\mu=0$, this Hamiltonian reduces to a pseudospin Heisenberg model which manifestly
exhibits the pseudospin SU(2) symmetry.

\section{Formalism}
The most general problem we would like to tackle, motivated by the issues discussed in the introduction,
is to understand the collective mode spectrum of a flowing superfluid. 
We begin by setting up a two-pronged approach to attack this problem. 
At small values of $U/t$, 
we build upon the mean field theory of the flowing superfluid to obtain the collective
mode spectrum using GRPA. At strong coupling, we
study the pseudospin model using Holstein-Primakoff spin wave theory. We find that
the GRPA results smoothly match on to the results from spin wave theory of the pseudospin
model as we increase $U/t$, showing that the GRPA correctly captures aspects of the strong 
coupling limit. This suggests that the GRPA may be a very good approximation to compute
the collective mode spectrum and address the breakdown of superflow over the
entire range of couplings.

\subsection{Generalized Random Phase Approximation for the Attractive Hubbard Model}
We begin by constructing the mean field theory of the uniform
superconducting state of the attractive Hubbard model in the presence of nonzero superflow.
The mean field Hamiltonian in the presence of a supercurrent is obtained by
forming Cooper pairs with nonzero momentum,
\bea
\label{eq:5}
H_{\rm MFT} &=& -t \sum_{\la i j \ra} (c_{i\sigma}^\dagger c^{\vphantom \dag}_{j\sigma} +
c_{j\sigma}^\dagger c^{\vphantom \dag}_{i\sigma}) - \mu\sum_{i\sigma}n_{i\sigma} \nonumber \\
&-& \Delta_0 \sum_{i} \left( {\rm e}^{i {\bf Q} \cdot {\bf r}_i} c^\dg_{i\upa}
c^{\dg}_{i\dna} + {\rm e}^{-i {\bf Q} \cdot {\bf r}_i} c^\pdg_{i\dna} c^\pdg_{i\upa}
 \right),
\eea
where we have set  $U \la c^\pdg_{i\dna} c^\pdg_{i\upa}\ra =
\Delta_0 {\rm e}^{i {\bf Q} \cdot {\bf r}_i}$ and absorbed the uniform Hartree shift into 
the chemical potential. 
In momentum space, the mean field Hamiltonian takes the form
\bea
\label{eq:6}
H_{\rm MFT} \!\!=\!\! \sum_{{\bf k}} \xi_{\bf k} c_{\bk \sigma}^\dg c^\pdg_{\bk \sigma}
\!\!-\!\! \Delta_0 \!\!\sum_{\bk}\! \left(\!c^\dg_{\bk \upa} c^\dg_{-\bk + \bQ \dna} \!+\!
c^\pdg_{-\bk + \bQ \dna} c^\pdg_{\bk \upa}\!\right),
\eea
where
$\xi_\bk \equiv-2t \epsilon_{\bk}-\mu$, with
$\epsilon_\bk \equiv \sum_{i=1}^{d} \cos(k_{i})$ ($d=2,3$ is the dimensionality of the lattice). 

We can diagonalize $H_{\rm MFT}$
by defining Bogoliubov quasiparticles (QPs), $\gamma$, via
\begin{eqnarray}
\label{eq:7}
 \left(\begin{array}{c}
 c_{\bk \uparrow} \\ c_{-\bk + \bQ \downarrow}^\dagger
 \end{array}\right)=\left(\begin{array}{cc}
u_\bk(\bQ) & v_\bk(\bQ) \\ - v_\bk(\bQ) & u_\bk(\bQ)
\end{array}\right)\left(\begin{array}{c} \gamma_{\bk \uparrow} \\ \gamma_{-\bk + \bQ \downarrow}^\dagger
\end{array}\right).
\end{eqnarray}
For simplicity of notation, we will refer to the Bogoliubov transformation coefficients
above as $u_\bk, v_\bk$ with the implicit understanding that they depend on $\bQ$.
Parametrizing 
$ u_\bk \equiv \cos(\theta_\bk)$, 
$v_\bk \equiv \sin(\theta_\bk)$,
and demanding that the transformed Hamiltonian be diagonal
leads to the condition
\be
\label{eq:8}
\tan(2\theta_\bk) = \frac {\Delta_0}{\frac{1}{2} (\xi_\bk + \xi_{-\bk + \bQ})}.
\ee
Defining
\be
\label{eq:9}
\Gamma_\bk = \sqrt{\frac{1}{4} (\xi_\bk + \xi_{-\bk + \bQ})^2 + \Delta^2_0},
\ee
we find that the Bogoliubov transformation coefficients must satisfy the relations
\bea
\label{eq:10}
u^2_\bk &=& \frac{1}{2} \left(1 + \frac{\xi_\bk + \xi_{-\bk + \bQ}}{2 \Gamma_\bk} \right), \nonumber \\
v^2_\bk &=& \frac{1}{2} \left(1 -  \frac{\xi_\bk + \xi_{-\bk + \bQ}}{2 \Gamma_\bk} \right), \nonumber  \\
u_\bk v_\bk  &=& \frac{\Delta_0}{2 \Gamma_\bk}.
\eea
In terms of the Bogoliubov QPs, the mean field Hamiltonian finally
takes the form
\begin{eqnarray}
\label{eq:11} 
H_{\rm MFT}=E_{\rm GS}+\sum_{\bk} E_\bk \gamma_{\bk \sigma}^\dg \gamma^\pdg_{\bk \sigma},
\end{eqnarray}
where $E_{\rm GS}$ denotes the ground state energy of $H_{\rm MFT}$
and $E_\bk$ denotes the Bogoliubov QP dispersion given by:
\bea
\label{eq:12}
E_\bk &=& \Gamma_\bk + \frac{1}{2} \left(\xi_\bk - \xi_{-\bk + \bQ} \right), \nonumber \\
E_{\rm GS} &=& \sum_\bk \left(\xi_\bk - E_\bk \right).
\eea
Demanding self-consistency of the mean field theory yields the gap and number
equations:
\begin{eqnarray}
\label{eq:13}
\frac{1}{U}&=& \frac{1}{N}\sum_{\bk} \frac{1}{2 \Gamma_\bk} (1 - n_{F}(E_{\bk}) - n_{F}(E_{-\bk + \bQ})), \nonumber \\
f&=&\frac{2}{N}\sum_{\bk} \left[u_\bk^2 n_{F}(E_{\bk}) + v_\bk^2 (1 - n_{F}(E_{-\bk + \bQ}))\right], \nonumber \\
\end{eqnarray}
where $f$ is the filling, i.e. the average number of fermions per site, 
and $N$ is the total number of sites. 
For given $U$, filling $f$ and flow momentum $\bQ$, these equations can be 
solved to obtain the SF order parameter $\Delta_0$ and the QP spectrum.

Going beyond mean field theory, we need to include fluctuations of the density
and the superfluid order parameter, which we will do within GRPA. 
We begin by considering perturbing external fields that couple to the density 
and order parameter modulation operators as:
\begin{eqnarray}
\label{eq:14}
H'_{\rm MFT} &=&H_{\rm MFT} -\!\! \sum_{i} [ h_{\rho}(i,t)\hat{\rho}_i \nonumber \\
&+& h_{\Delta}(i,t)\hat{\Delta}_i e^{i \bQ \cdot \br_i} + h_{\Delta}^{*}(i,t)\hat{\Delta}_i^\dagger 
e^{- i \bQ  \cdot \br_i}],
\end{eqnarray}
where
\begin{eqnarray}
\label{eq:15}
\hat{\rho}_i &=& \frac{1}{2} c_{i\sigma}^\dagger c^\pdg_{i\sigma}, \nonumber \\
\hat{\Delta}_i &=& c_{i\downarrow} c_{i\uparrow}.
\end{eqnarray}
Going to momentum space,
\begin{eqnarray}
\label{eq:16}
H'_{\rm MFT}=H_{MFT} - \frac{1}{N}
\sum_{\bK}\,\, h_\alpha(\bK,t) \hat{O}^\dg_{\alpha}(\bK),
\end{eqnarray}
where $\alpha = 1,2,3$, summation over $\alpha$ is implicit, and $\hat{\bf O}^\dg(\bK) \equiv \{\hat{\rho}^\pdg_{-\bK}, \hat{\Delta}^\pdg_{-\bK + \bQ},
\hat{\Delta}_{\bK + \bQ}^\dagger\}$
is the vector of fermion bilinear operators corresponding to density and superfluid order parameters
at nonzero momenta, with the order parameters given by:
\begin{eqnarray}
\label{eq:17}
\hat{\rho}^\pdg_{\bK}&\equiv& \frac{1}{2}\sum_{\bk} c_{\bk \sigma}^\dagger c^\pdg_{\bk + \bK \sigma}, \nonumber  \\
\hat{\Delta}^\pdg_{\bK}&\equiv& \sum_{\bk} c^\pdg_{-\bk + \bK \downarrow} c^\pdg_{\bk \uparrow}, \nonumber \\
\hat{\Delta}^\dg_{\bK}&\equiv& \sum_{\bk} c^\dg_{\bk \upa} c^\dg_{-\bk + \bK \dna}.
\end{eqnarray}
The perturbing fields correspond to
\bea
\label{eq:18}
h_1(\bK,t)&=&h_\rho(\bK,t), \nonumber \\
h_2(\bK,t)&=&h_\Delta(\bK,t), \nonumber \\
h_3(\bK,t)&=&h_\Delta^*(-\bK,t).
\eea
We define the susceptibility matrix as
\be
\label{eq:19}
 \la\hat{O}_{\alpha}(\bK,\omega)\ra = \chi_{\alpha\beta}^{0}(\bK,\omega) h_{\beta}(\bK,\omega).
\ee
The derivation of this bare susceptibility matrix as well as the explicit expressions for
the elements of this matrix are given in Appendix B.

\begin{figure}
\includegraphics[width=2.8in]{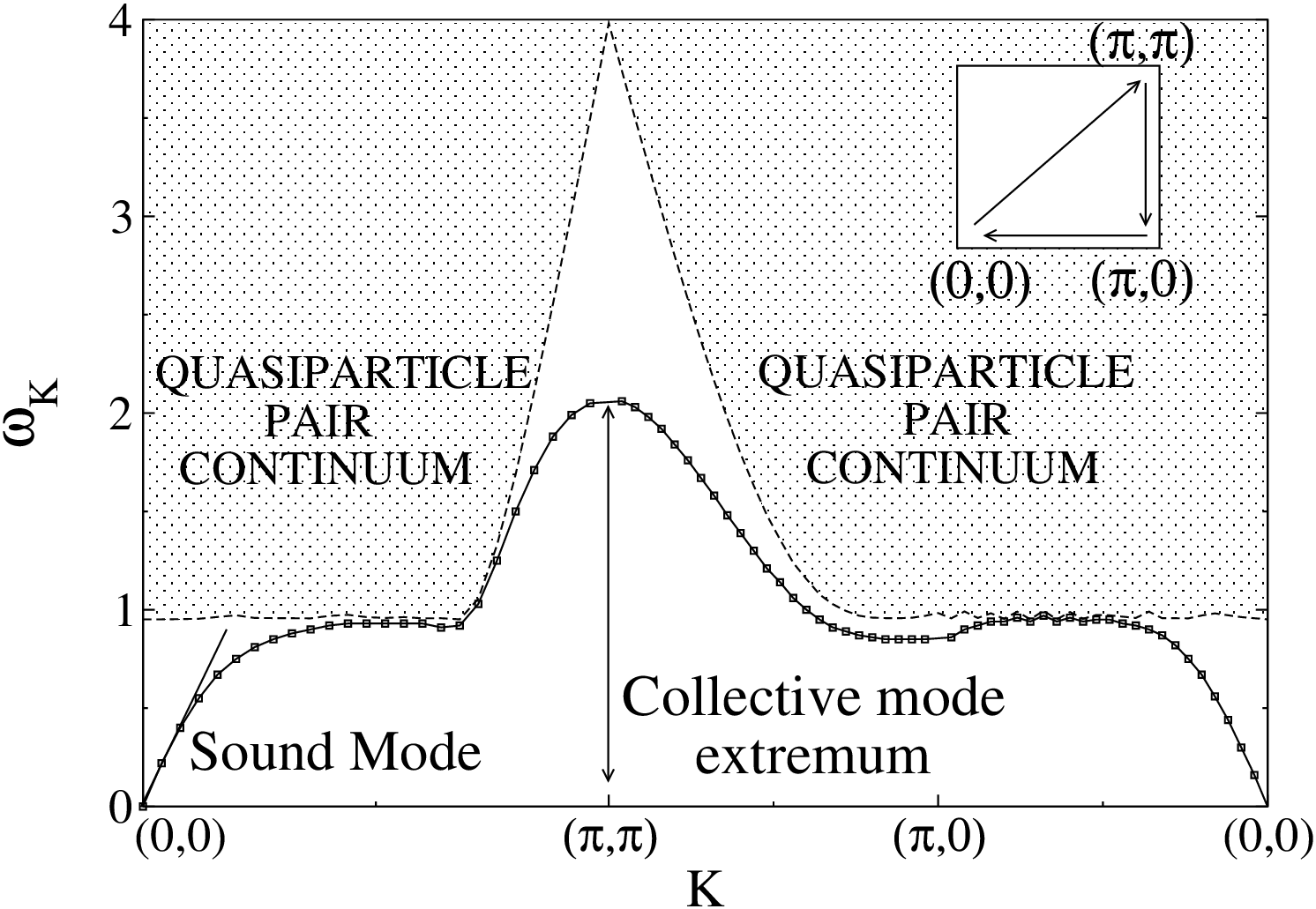}
\caption{Illustrative example of the collective mode dispersion and quasiparticle-pair 
continuum in 2D, for zero superflow ($\bQ=0$) with $U/t=3$ and $f=0.2$ fermions per site,
along the contour displayed in the inset.}
\label{fig:rpaexample}
\end{figure}

In order to account for interaction effects at the GRPA level, we note that the interaction 
effectively induces internal fields which renormalize the applied external field so that the
susceptibility within the GRPA can be expressed as:
\be
\label{eq:20}
\chi^{GRPA}_{\alpha\beta}=(1-U \chi^{0} D)^{-1}_{\alpha\nu}\chi^{0}_{\nu\beta}. 
\ee
Here, $D={\rm diag} \{ 2,1,1 \}$ is a diagonal
matrix that encodes the decoupling of the interaction Hamiltonian. The derivation
of this expression in explained in Appendix B.
This GRPA susceptibility will diverge when the determinant ${\rm Det}(1-U\chi^0 D)$
becomes zero (or equivalently, one of the eigenvalues of this matrix vanishes). We
identify the locus of real frequencies, $\omega\equiv \omega(\bK)$, where this happens, as
the dispersion of a sharp (undamped) collective mode.

Fig.~\ref{fig:rpaexample} provides an illustrative example of the collective mode spectrum 
obtained using the GRPA in 2D with $U/t=3$ and a filling $f=0.2$ fermions per site in
the absence of superflow ($Q=0$). We find a linearly dispersing superfluid 
``phonon'' mode
at small momenta and low energy. The slope of this linear dispersion is the sound velocity. 
The collective mode disperses as a function of momentum
over the Brillouin zone and exhibits an extremum at the edges of the Brillouin zone. In 
the example here, it is a maximum at $\bK=(\pi,\pi)$, but with increasing interaction strength
and at fillings closer to $f=1$, the spectrum exhibits a minimum at $\bK=(\pi,\pi)$ arising
from strong short distance density correlations.
At high energies, there is an onset of a two-quasiparticle continuum
where the collective mode can decay by creating two Bogoliubov quasiparticles with
opposite spins in a manner which conserves energy and momentum.
Once the collective mode energy
goes above the lower edge of the two-particle continuum of Bogoliubov QP excitations,
it will cease to be a sharp excitation and acquire a finite lifetime. The collective mode energy as well as the
pair continuum change in the presence of superflow as discussed subsequently.

\subsection{Strong Coupling Limit: Spin Wave Analysis of Pseudospin Model}

In the pseudospin model, a state with
nonzero supercurrent is obtained by imposing a phase twist on the non-flowing
ground state $\vert 0\ra $ as 
\be
\label{eq:21}
\vert \bQ \ra = \exp\left[-i \sum_i T^z_i \bQ \cdot  \br_i\right]\vert 0\ra.
\ee
Equivalently, we can make a unitary transformation to work with the Hamiltonian
\bea
\label{eq:22}
\nonumber H_{\rm eff} (\bQ) &=& J \sum_{\la ij \ra}[T_{i}^{z} T_{j}^{z} + 
(T_i^x T_j^x +T_i^y T_j^y)\cos(\bQ \cdot \br_{ij}) \\
&-& (T_i^x T_j^y \!-\! T_i^y T_j^x) \sin(\bQ \cdot \br_{ij})]
 - \mu \sum_{i} T^z_i, \nonumber \\
\eea
where $\br_{ij} \equiv \br_i - \br_j$. This amounts
to transforming to a reference frame where the superfluid is at rest.

The classical ground state $|0\ra_c$,
which is 
the ground state obtained under the assumption that the pseudospins are classical vectors, 
is equivalent to the mean-field ground state of the interacting fermion model and
can be parametrized by specifying the classical
vector $\bT^c_i$ at each lattice site. This is given by
\begin{equation}
\label{eq:23}
\bT^c_i \equiv S (\eta_i \sin\theta, 0, \cos\theta),
\end{equation}
where the pseudospin magnitude $S=1/2$. 
The angle $\theta$ is related to the filling, $f$, defined earlier, as
\be
\label{eq:24}
f -1 = \cos\theta.
\ee
Working at fixed filling, the chemical potential $\mu$ is given by
\begin{equation}
\label{eq:25}
\mu = 2 J S \cos(\theta) (\epsilon_0 + \epsilon_\bQ),
\end{equation}
where $\epsilon_\bQ \equiv \sum_{i=1}^{d} \cos(Q_{i})$ as before.

\begin{figure}
\includegraphics[width=2.8in]{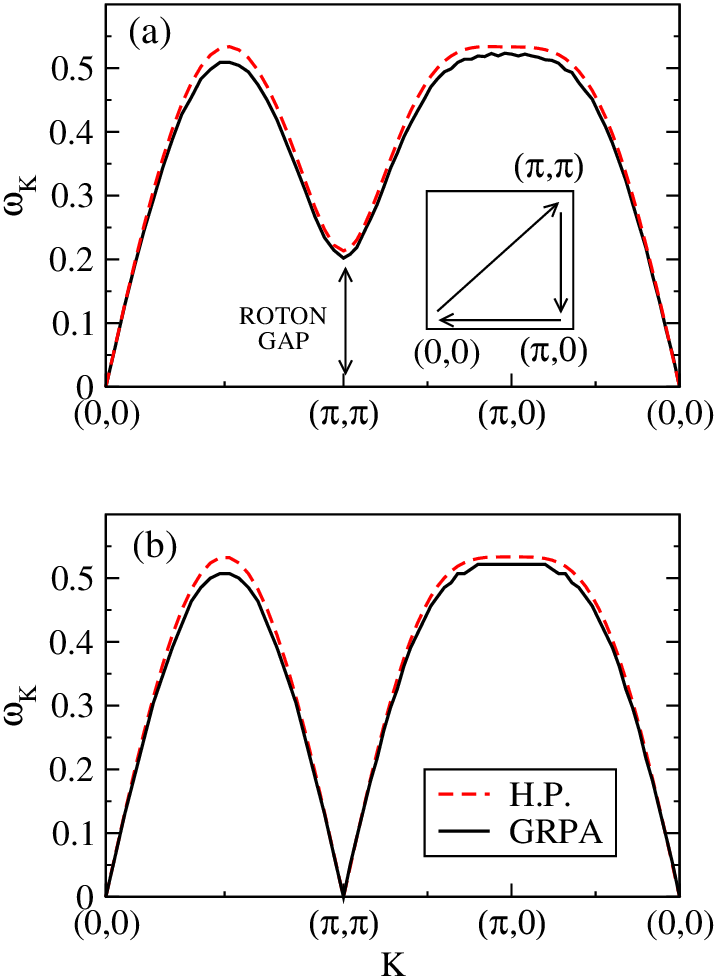}
\caption{(Color online) Collective mode energy at zero superflow
in 2D at strong coupling, $U/t=15$. The dispersion is shown along the indicated
contour in the Brillouin zone for different fillings: (a) $f=0.8$ fermions
per site and (b) $f=1.0$ per site. The GRPA result (solid line) is in good agreement
with the Holstein-Primakoff spin wave result (dashed line, HP) for the strong coupling pseudospin
model. The roton minimum has a small gap at $f=0.8$ but becomes a gapless mode at
$f=1.0$ due to the pseudospin SU(2) symmetry discussed in the text.}
\label{fig:hprpa}
\end{figure}

We compute the excitations about the ground state $\vert 0\ra_c$ for the Hamiltonian
$H_{\rm eff}(\bQ)$ using a Holstein-Primakoff (HP)
approach~\cite{auerbach} by rotating to new pseudospin operators $\tilde{T}$ given by
\begin{eqnarray} 
\label{eq:26}
 \nonumber \tilde{T}_i^z &=& T_i^z \cos(\theta) + \eta_i T_i^x \sin(\theta), \\
 \nonumber \tilde{T}_i^x &=& -T_i^z \sin(\theta) + \eta_i T_i^x \cos(\theta), \\
\tilde{T}_i^y &=& \eta_i T_i^y,
\end{eqnarray}
and expressing $\tilde{T}$ in terms of HP bosons.
Expanding the Hamiltonian to ${\cal O}(S)$, we find
\begin{eqnarray}
\label{eq:27}
H&=& E_{\rm c} + \delta E_{\rm q} 
+ \sum_{\bK} \omega_\bK (\bQ) b^\dagger_\bK b^\pdg_\bK,
\end{eqnarray}
where
\bea
\label{eq:28}
E_{\rm c} \!\!&\!\!=\!\!&\!-NJS^2 [\cos^2\theta\epsilon_{0} +(1+\cos^2\theta)\epsilon_{\bQ} ], \nonumber \\
\delta E_{\rm q} \!\!&\!\!=\!\!&\!\frac{JS}{2}\!\sum_{\bK} [\epsilon_{\bK}\! \sin^2\!\theta \!-\! 
\frac{
(1\!\!+\!\!\cos^2\!\theta)}{2}\!
(\epsilon_{\bK\!+\!\bQ}\!\!+\!\!\epsilon_{\bK\!-\!\bQ})],
\eea
represent, respectively, the classical ground state energy and the leading quantum correction
to the ground state energy. 
The spin-wave dispersion $\omega_\bK (\bQ)$ is given by
\bea
\label{eq:29}
\omega_\bK (\bQ)&=& 2JS (\beta_\bK (\bQ)+\sqrt{\alpha_\bK^2 (\bQ)-\gamma_\bK^2 (\bQ)}),
\eea
with
\bea
\label{eq:30}
\alpha_\bK (\bQ)\!\!&=&\!\! \!\epsilon_\bQ\!+\!\frac{\sin^2 \theta}{2}\epsilon_\bK\!-\!\frac{(1\!+\!\cos^2 \theta)}{2}
\!\left(\!\!\frac{\epsilon_{\bK + \bQ}\!+\!\epsilon_{\bK - \bQ}}{2}\!\!\right)\!, \nonumber  \\
\beta_\bK (\bQ)&=&\frac{1}{2} \cos\theta\left(\epsilon_{\bK - \bQ} - \epsilon_{\bK + \bQ}\right), \nonumber \\
\gamma_\bK (\bQ)&=&\frac{1}{4} \sin^2\theta \left(2 \epsilon_\bK  + 
\epsilon_{\bK + \bQ} + \epsilon_{\bK - \bQ} \right).
\eea

An illustration of the collective mode dispersions obtained using this approach is
shown in Fig.~\ref{fig:hprpa} at a strong coupling value of the interaction $U/t=15$
for two different fillings, $f=0.8,1.0$ fermions per site.
We find that these dispersions are in good agreement with those obtained using the GRPA
which serves to show that the GRPA also correctly captures aspects of the strong coupling
limit. As seen from the dispersion, there is a linearly dispersing superfluid 
``phonon'' mode
at small $K$. In addition, for $f=0.8$, we find a low energy
``roton minimum'' at $K=(\pi,\pi)$ which arises from strong local pair density correlations in the
superfluid ground state. For $f=1.0$, the gapless roton mode at $K=(\pi,\pi)$ is a manifestation
of the pseudospin SU(2) degeneracy, discussed earlier, between the crystalline
CDW ground state and the uniform superfluid ground state. 

\section{Collective Modes in the Stationary Superfluid}

\begin{figure}
\begin{center}
\includegraphics[width=2.8in]{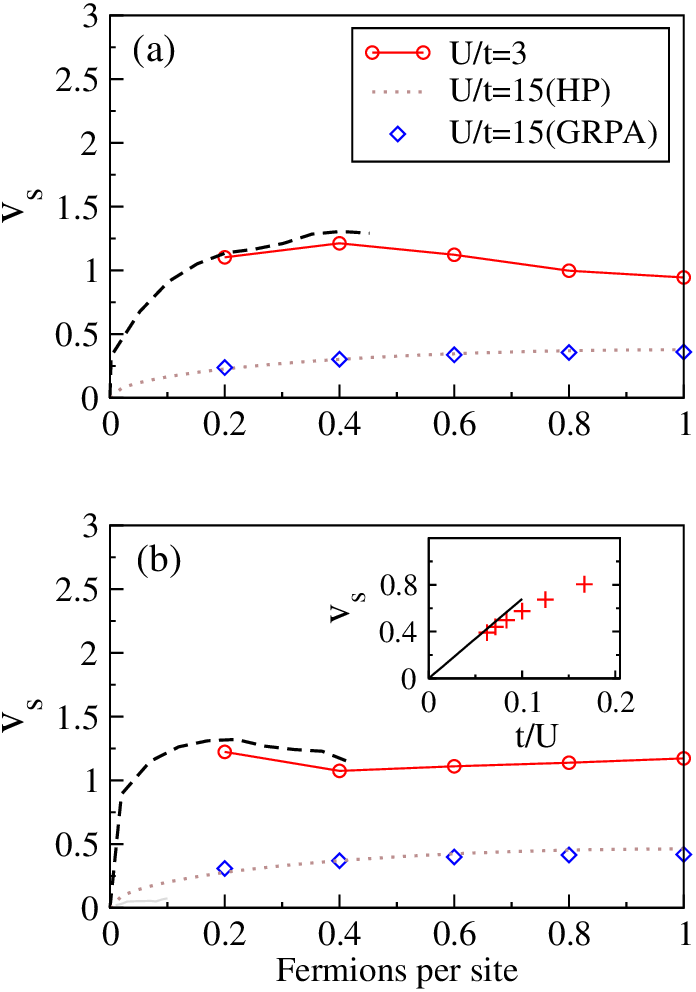}
\caption{(Color online) The sound mode velocity, $v_s$, as a function of the fermion filling $f$
in (a) 2D and (b) 3D for $U/t=3$ (circles) and $U/t=15$ (diamonds). Solid line is a guide to the eye.
The dashed lines are the weak coupling result,
$v_s = (v_F /\sqrt{d})[1-UN(0)]^{1/2}$, from Ref.[\onlinecite{Randeria94}] for $U/t=3$, with
$N(0)$ being the non-interacting density of states (per spin) at the Fermi level.
The dotted line indicates the Holstein Primakoff spin-wave result for $U/t=15$.
The inset to (b) shows the expected $t/U$ scaling of $v_s/t$ for $U/t \gg 1$ in 3D.}
\label{fig:vsound}
\end{center}
\end{figure}

\begin{figure}
\begin{center}
\includegraphics[width=2.8in]{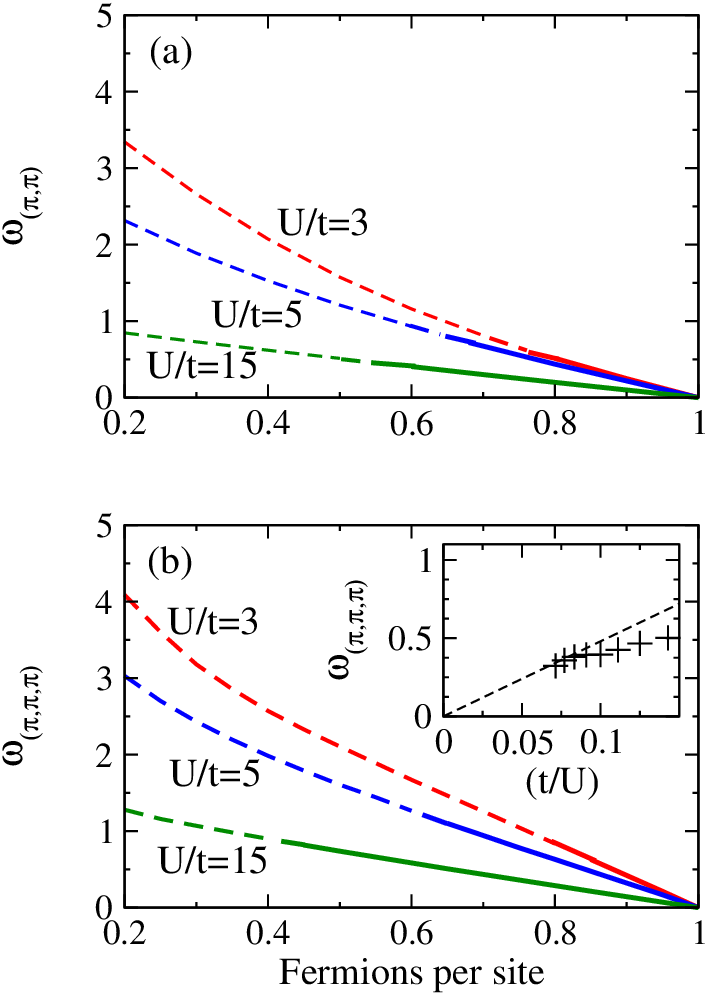}
\caption{(Color online) The energy of the collective mode at (a) $(\pi,\pi)$ in 2D and (b) $(\pi,\pi,\pi)$ in
3D 
for different interaction strengths. The dashed (solid) lines indicate  
that the mode energy corresponds to a local maximum (minimum) of the dispersion. 
The inset shows the collective mode energy in 3D at $(\pi,\pi,\pi)$ at a filling of $f=0.8$ fermions per site, 
as a function of $t/U$. Inset shows a comparison of the GRPA result (crosses) 
with the strong coupling spin-wave theory
result (dashed line).}
\label{fig:Eroton}
\end{center}
\end{figure}

We have seen that the collective mode spectrum obtained using the GRPA or
the strong coupling analysis has a superfluid phonon mode at small momenta
and a roton minimum at $\bK=(\pi,\pi)$ in 2D or at $\bK=(\pi,\pi,\pi)$ in 3D. The 
phonon mode is a reflection of the fact that the superfluid ground state breaks
a global U(1) symmetry -- it is thus the long wavelength Goldstone mode
associated with this symmetry breaking.
The roton mode arises from strong local density correlations,
analogous to the roton mode in $^4$He. However, unlike in $^4$He, the rotons
only exist at this commensurate wave vector on the lattice
and do not form an incommensurate
ring of wave vectors. Further, the roton mode becomes gapless at $\mu=0$ and 
this gaplessness arises from a pseudospin SU(2) symmetry which leads to a
degeneracy between the superfluid and CDW ground states as
discussed earlier.

\subsection{Sound Velocity}
Fig.~\ref{fig:vsound} shows the sound mode velocity $v_s$, extracted using the GRPA,
over a range of densities and interaction
strengths in 2D and 3D. In the limit of low filling and weak interaction,
our calculations of $v_s$ agree with the results of Belkhir and Randeria,~\cite{Randeria94}
although strong finite size effects and a small pairing gap limit our numerical
exploration of the very weak coupling regime $U/t\ll 1$. 
At strong coupling, $U/t \gg 1$, our results from the GRPA are in good agreement with the
spin wave result $v_s= (4t^2/U) \sqrt{d} \sqrt{2f-f^2}$. The linear scaling of $v_s/t$
with $t/U$, expected for $U/t \gg 1$, is illustrated in the inset of Fig.~\ref{fig:vsound}(b).

\subsection{Roton Gap}
Fig.~\ref{fig:Eroton} shows the energy of the collective mode at $K=(\pi,\pi)$ in 2D and $K=(\pi,\pi,\pi)$ in 3D. 
For low fillings and weak to intermediate coupling strengths, the
mode energy shows a maximum at this momentum. For strong enough interactions, or for
fillings near $f=1.0$, however, the mode energy exhibits a local minimum and can
be justifiably identified as a roton mode. The roton gap clearly scales as $t/U$ for
$U/t \gg 1$ as seen from the inset of Fig.~\ref{fig:Eroton}b. Further, the roton energy
goes to zero linearly as $f\to 1$. This is due to the chemical potential $\mu$, which tunes the
filling $f$ away from half-filling, and explicitly breaks the pseudospin SU(2) symmetry 
present at $f=1$.

\section{Superflow Instabilities}

Superflow instabilities are easily understood in the context of bosonic superfluids possessing Galilean invariance. Superflow can then be simply thought of as a Galilean boost performed on a stationary SF. The excitations in the flowing frame are the same as that of the stationary SF, except their energies undergo a Doppler shift. At a certain critical flow velocity, the Doppler shift renders the excitations gapless at some momentum. These gapless excitations are then populated, flow energy is transferred to these bosonic excitations, and superfluidity is lost. The critical velocity is given by $v_{\rm crit} = {\rm min}(\omega_q/q)$, where $\omega_q$ is the energy of an excitation of the stationary SF at momentum q.

Our system, a paired fermionic superfluid on a lattice, 
is much richer. The excitations are fermionic Bogoliubov quasiparticles and a bosonic collective mode,
corresponding to coupled
quantum-mechanical fluctuations of the superfluid order parameter and the density. 
The effect of flow is now more than simply introducing a Doppler shift in excitation energies since the system explicitly breaks Galilean invariance. 
The SF order parameter and the quasiparticle dispersion of the mean field theory are strongly renormalized. In addition, the collective mode dispersion also changes strongly. This is easily seen in the strong coupling 
pseudospin model, where the dispersion of spin wave collective modes is given by Eq.(\ref{eq:29}). 
The term $\beta_\bK (\bQ)$ in this equation is the Doppler shift and clearly depends on the flow momentum. 
However, the terms $\alpha_\bK (\bQ)$ and $\gamma_\bK (\bQ)$ are strong functions of the flow momentum too. Further, on the lattice as opposed to the continuum, 
the Doppler shift vanishes at momentum points corresponding to the Brillouin zone edges. The
only effect of superflow in this case is to cause a strong dispersion renormalization via
its effect, at strong coupling, on $\alpha_\bK(\bQ),\gamma_\bK(\bQ)$.

Due to these effects of imposed superflow, our system undergoes three broad kinds of instabilities, which we call ``depairing", ``Landau" and ``dynamical".
Below, we describe each of these and map out a stability ``phase diagram" as a function of dimensionality, interaction strength and density, indicating the critical flow momentum beyond which the uniform
flowing superfluid is unstable. Here we will restrict our attention to superflow momenta
$\bQ=Q_x\hat{x}$.

\subsection{Depairing Instability}
The system undergoes a depairing instability when the self-consistently calculated SF order parameter, $\Delta$, vanishes in the mean field theory of the flowing SF. At the critical flow momentum, the energy cost of  flow outweighs the condensation energy gain, and the system goes into an unpaired 
normal state. 

This instability is close to, but not identical with, the quasiparticles becoming gapless due to the flow-induced Doppler shift. In 2D, we find that these two phenomena occur at the same value of the flow momentum. 
In 3D however, superfluidity persists beyond the point where quasiparticles become gapless. There is a small window of {\em gapless superfluidity},  
where negative energy Bogoliubov quasiparticle states are occuppied. This is analogous to what has been claimed to occur, for instance, in superfluid
$^3$He in the presence of superflow.~\cite{volovik}
In this case, although there are negative energy quasiparticle excitations, the system cannot
arbitrarily lower its energy by occupying such states since the Bogoliubov excitations are
fermionic. The gapless superfluid thus survives as a stable intermediate phase in 3D.~\cite{Wei09}

\subsection{Landau Instability}

\begin{figure}
\begin{center}
\includegraphics[width=2.8in]{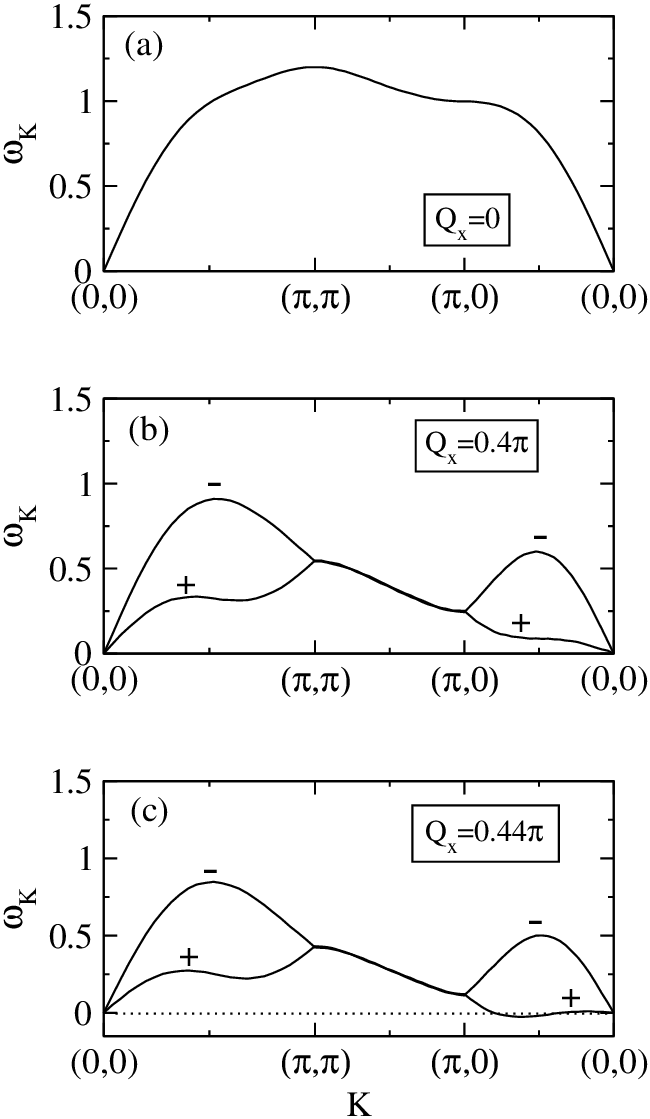}
\caption{Landau instability - Collective mode spectrum (from GRPA) in 2D with $U/t=7$ at a filling of 
$f=0.4$ fermions per site. Collective mode dispersion is 
shown at (a) zero flow, (b) just below, and (c) just above, the critical flow momentum for the
Landau instability. The $+ (-)$ sign indicates the mode is at wavevector $+\bK (-\bK)$ which has an
$x$-component
along (opposite to) the flow direction so that it is Doppler shifted down (up) in energy.
The collective mode frequency becomes negative at an incommensurate wavevector.}
\label{Landau}
\end{center}
\end{figure}

A Landau instability occurs when the collective mode energy hits zero and becomes negative, as shown in Fig.~\ref{Landau}. In the strong coupling limit, with the collective mode dispersion given
in Eq.(29), this case is described by $\alpha_\bK(\bQ) \geq \gamma_\bK(\bQ)$ and $\beta_\bK(\bQ) < 
-\sqrt{\alpha_\bK ^2(\bQ) + \gamma_\bK^2 (\bQ)}$. In the GRPA approach, we find that the renormalized susceptibility, $\chi^{GRPA}$, diverges for two real negative frequencies. These frequencies match smoothly onto the results of the spin wave frequencies at strong coupling.

As discussed in Ref.[\onlinecite{Burkov08}], Landau instability is not an instability of linearized dynamics of small fluctuations around the 
uniformly flowing state, since some form of mode coupling
is necessary to transfer the energy of the superflow into these negative-energy modes.  Its full theoretical description is thus rather complicated. 
Here we restrict ourselves to only finding the critical flow momentum for this instability. 

We find that the Landau instability occurs either at small momenta, corresponding to the dispersion of the sound mode going below zero, or at some finite incommensurate momentum. In the case of low filling, unless preempted by depairing, we see a Landau instability of the sound mode. For moderate values of U and the filling, we see Landau instabilities at large incommensurate momenta (as in Fig.~\ref{Landau}). As the filling is reduced, this incommensurate 
wavevector moves towards the Brillouin zone centre, so that in the low density limit, it is the long wavelength sound mode that becomes unstable. Similar incommensurate instabilities have also been found in Ref.[\onlinecite{danshita}].

\begin{figure}
\begin{center}
\includegraphics[width=2.8in]{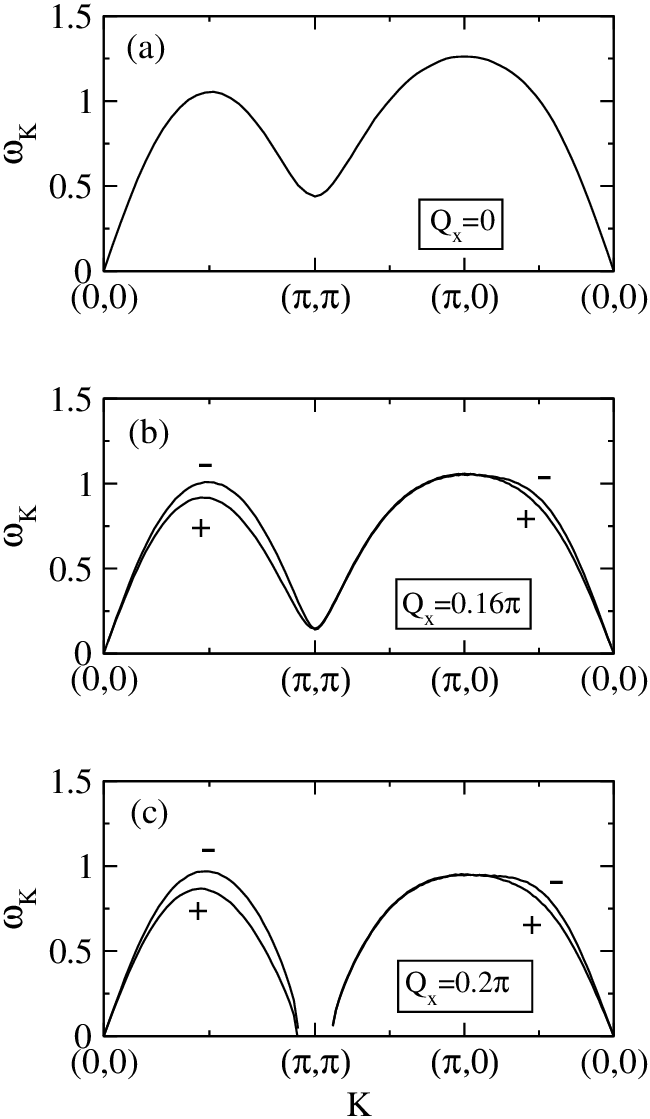}
\caption{Dynamical Instability (Commensurate)- Collective mode spectrum (from GRPA) for $U/t=5$, filling $f=0.8$ fermions per site on a 2D square lattice. 
The $+ (-)$ sign indicates the mode is at wavevector $+\bK (-\bK)$ which has an
$x$-component
along (opposite to) the flow direction.
As the flow momentum Q is increased, the collective mode frequency at $\bK=(\pi,\pi)$ decreases until it hits zero and becomes complex. 
This gives rise to a dynamical instability associated with the ``checkerboard" 
CDW order. The part of the dispersion, around $(\pi,\pi)$, which corresponds to unstable modes is not shown.}
\label{dyncomm}
\end{center}
\end{figure}

\subsection{Dynamical Instability}

The underlying lattice potential allows for a new possibility for the 
breakdown of superflow, namely through dynamical instabilities.~\cite{Niu01}
This corresponds to the eigenfrequency of the collective mode dispersion in our system becoming zero, and subsequently becoming complex.

The concept of dynamical instability is particularly simple to understand for weakly interacting bosons.
In this case, the instability coincides with the point where 
the effective mass of the bosons changes sign as a function of the superflow momentum,~\cite{Niu01} leading to 
runaway growth of phase and density fluctuations which eventually
destroy superfluidity.

For strongly interacting bosons the situation is more interesting. 
It has been shown~\cite{Altman05} that with increasing interaction strength at a commensurate
filling (integer number of bosons per site), the dynamical instability occurs at a smaller
and smaller superflow momentum. The critical flow momentum eventually tends to zero at the equilibrium superfluid to
Mott insulator transition, scaling as the inverse of the divergent correlation length
associated with this
Mott transition.

Here, in the fermionic analog of the problem of the critical superflow in a strongly correlated superfluid, 
we find, in addition, 
an entirely different kind of dynamical instability, associated with emergent density-wave 
order for a large range of fillings, wherein the mode energies become complex at nonzero wavevectors.
In the strong coupling limit, this happens when $|\alpha_\bK (\bQ)|$ becomes smaller than $|\gamma_\bK (\bQ)|$. In the GRPA approach, at a dynamical instability, we find no real frequency at which any eigenvalue
of the inverse GRPA susceptibility vanishes.
The GRPA results for the critical flow momentum $\bQ$, and the wavevector $\bK$ 
at which the dynamical instability occurs, match smoothly onto the spin wave results at strong coupling. 
In this limit it is easy to check analytically that
the dynamical instability associated with this emergent charge order is not preempted by a Landau instability.
The appearance of complex collective mode frequencies indicates an
exponential growth of density fluctuations around the uniformly flowing state, at a wavevector $\bK$.

\begin{figure}
\begin{center}
\includegraphics[width=2.8in]{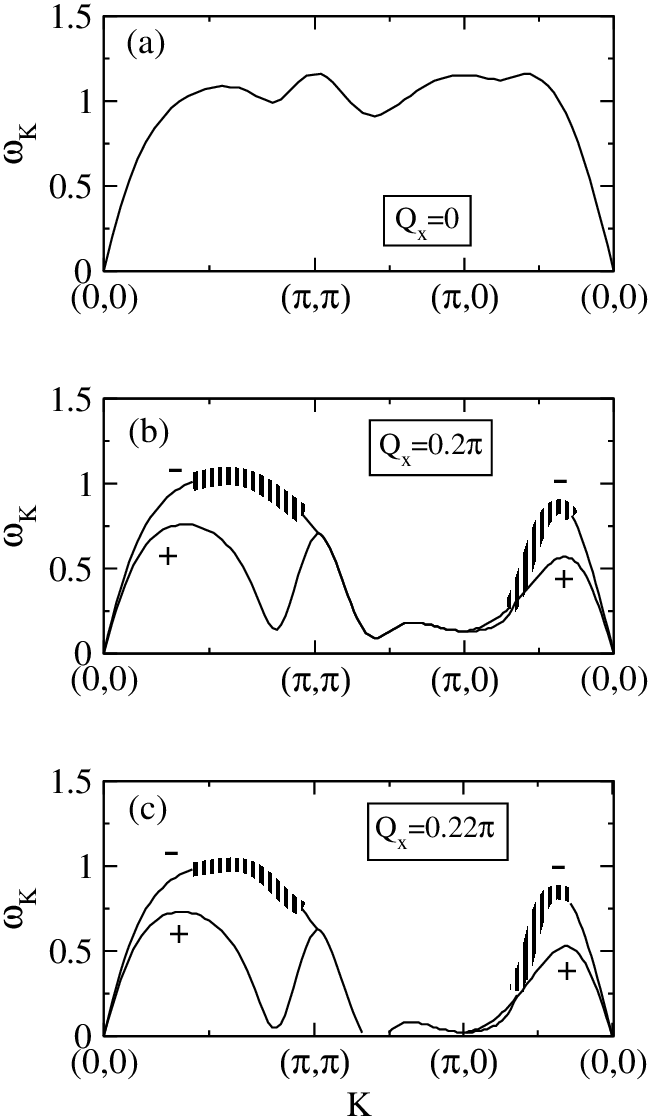}\\
\caption{Dynamical Instability (Incommensurate) - Collective mode spectrum (from GRPA) for $U/t=3$, filling $f=0.6$ fermions per site on a 2D square lattice. 
The $+ (-)$ sign indicates the mode is at wavevector $+\bK (-\bK)$ which has an
$x$-component
along (opposite to) the flow direction.
As the flow momentum $Q_x$ is increased beyond $0.2\pi$, the collective mode frequency becomes complex at an incommensurate wavevector. The complex frequencies in the unstable region are not shown in the figure.}
\label{dynincomm}
\end{center}
\end{figure}

We find two kinds of such dynamical instabilities --- associated with commensurate (Fig. \ref{dyncomm}) or
incommensurate (Fig. \ref{dynincomm}) charge order. 
We see the former for a large range of fillings in the vicinity of  $f=1$, for all interaction strengths. The wavevector 
at which this instability happens is at the  Brillouin zone 
corner - $(\pi,\pi)$ in 2D and $(\pi,\pi,\pi)$ in 3D. This corresponds to an instability towards a ``checkerboard" CDW state, 
with a density modulation of opposite sign on the two sublattices of the square or cubic lattice.
The incommensurate dynamical instability, on the other hand, occurs at wavevectors corresponding to various incommensurate ordering patterns and it arises as follows.
In the presence of superfow, the dispersion of the Bogoliubov quasiparticles is renormalized and Doppler
shifted, giving rise to multiple minima separated by a nonzero wavevector.
This leads to a peak in the
bare mean field susceptibility, $\chi^0$ at this momentum. 
The interaction renormalizes this peak into a divergence and leads to a finite momentum pairing instability
of Bogoliubov quasiparticles. 
This dynamical incommensurate instability is thus a non-trivial, interaction-driven phenomenon, which only 
occurs at intermediate coupling strength.
This phenomenon is somewhat analogous to the formation of excitonic condensates in indirect band-gap 
semiconductors.~\cite{Halperin1968}

\subsection{Stability phase diagrams}
\begin{figure}
\begin{center}
\includegraphics[width=2.8in]{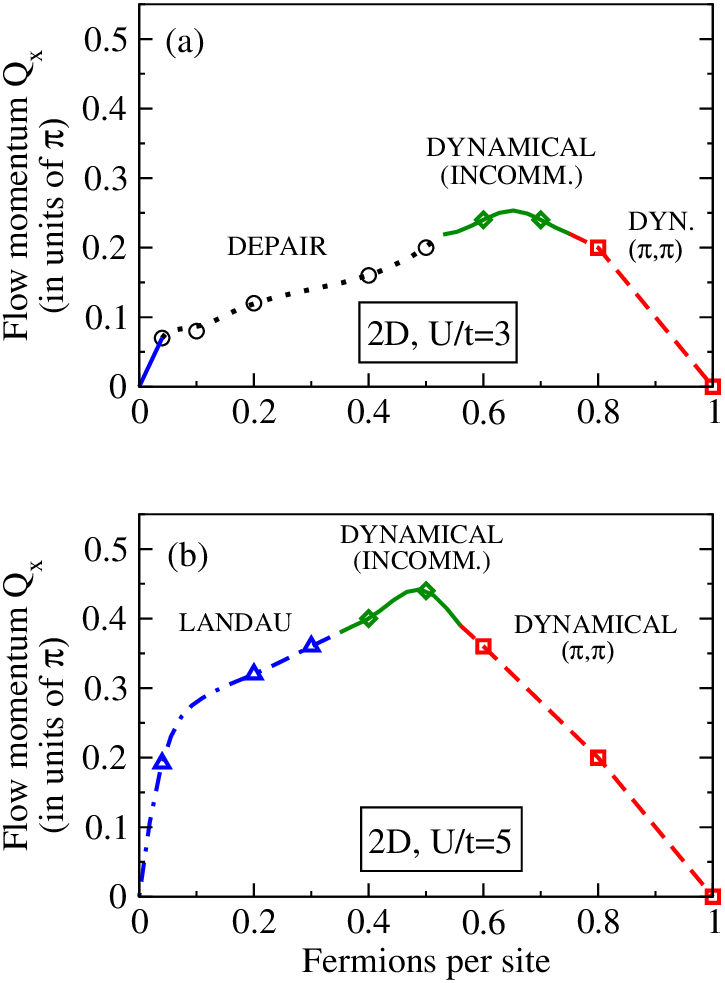}
\caption{(Color online) Stability phase diagram for a 2D square lattice case with (a) $U/t=3$ and (b) $U/t=5$. For every filling, the plot shows the first instability that is encountered as the flow is increased. The solid (blue) line in the low density limit in (a) indicates the region where we expect to see a Landau instability, but finite size effects prevent us from accessing the area.
The other transitions in the figure correspond to Depairing (circles, dotted line), Incommensurate Dynamical Instability
(diamonds, solid line), Commensurate Dynamical Instability (squares, dashed line), Landau Instability (triangles, 
dash-dotted line). At very large interaction, the diagram looks similar to $U/t=5$, except that the incommensurate 
dynamical instabilities disappear.}
\label{2dpd}
\end{center}
\end{figure}

\begin{figure}
\begin{center}
\includegraphics[width=2.8in]{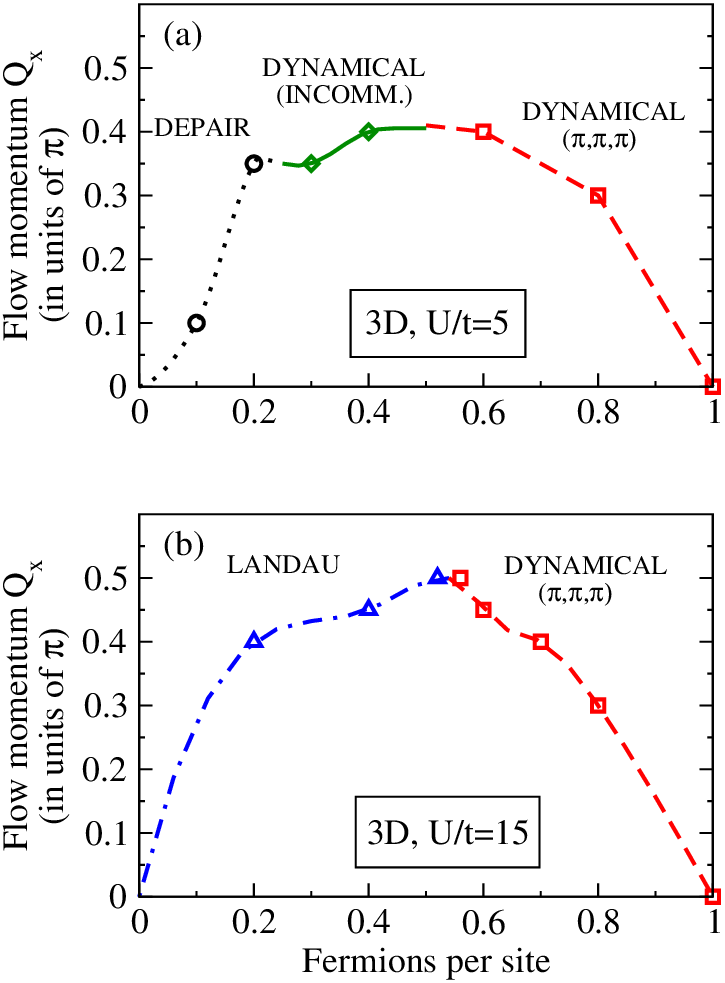}
\caption{(Color online)
Stability phase diagram for a 3D cubic lattice with (a) $U/t=5$ and (b)$U/t=15$. For each filling, the instability first encountered as the flow is increased is shown.
The transitions in the figure correspond to Depairing (circles, dotted line), Incommensurate Dynamical Instability
(diamonds, solid line), Commensurate Dynamical Instability (squares, dashed line), Landau Instability (triangles, 
dash-dotted line).}
\label{3dpd}
\end{center}
\end{figure}

Taking these instabilities into account, we map out superflow stability phase diagrams in 2D (Fig. \ref{2dpd}) and 3D (Fig. \ref{3dpd}). 
These plots show the first instability that is encountered as imposed flow is increased, for different values of filling. Values of $U$ for the plots 
have been chosen so as to illustrate all possibilities. We see that the commensurate dynamical ``checkerboard" CDW  instability comes into play around 
$f=1/2$ for all values of the interaction strength and is the dominant instability all the way to $f=1$, where the critical flow momentum vanishes, 
reflecting the degeneracy between the SF and CDW states. 

In the low density limit, the system is similar to a continuum Fermi gas. In 3D, the density of states at the Fermi level vanishes. 
At low interaction strength, this leads to the pairing gap $\Delta$ being exponentially suppressed. The sound velocity, on the other hand, is proportional to the Fermi velocity, so that
$v_s \sim f^{1/3}$. This leads to a rather sharp drop of the sound velocity as $f \to 0$ but the
gap drops to zero much faster.  
Therefore, in the 3D case, at weak interaction and low filling, a small imposed flow will drive $\Delta$ to zero before the flow velocity 
exceeds the sound velocity, leading to a depairing instability as can be be seen in Fig. \ref{3dpd}.

In contrast, in the low density continuum limit in 2D, 
the density of states goes to a constant, which means the pairing gap stays finite 
as $f \rightarrow 0$. With imposed flow, the first instability that one encounters is then the Landau instability, which will 
happen when the flow velocity exceeds the sound velocity which scales as $v_s \sim f^{1/2}$.
We have not been able to numerically uncover the Landau instability in this regime
due to severe finite size effects.

At intermediate densities, and at intermediate values of the interaction strength, the leading instability is the incommensurate dynamical 
instability, corresponding to the emergence of an incommensurate CDW, characterized by a wavevector which depends on the 
density and the interaction strength. 
This instability does not happen in either the strong-coupling limit, where the Anderson pseudospin description is appropriate,~\cite{Burkov08}
nor in the weak-coupling BCS superfluid limit, where the instability at intermediate densities is due to depairing. 
This dynamical instability is a nontrivial intermediate-coupling phenomenon and is one of the most interesting results reported 
in this work. 

An interesting question that arises in relation to the emergent CDW-driven dynamical instabilities, is the fate of the system 
beyond the critical flow momentum. 
We show, in Appendix C, that beyond the commensurate CDW dynamical 
instability threshold the system behaves as a ``flowing supersolid'' at the mean field level. However, we argue
that fluctuations beyond
mean field theory are expected to destroy this flowing supersolid. An explicit example 
of this,
in the strong coupling limit, has been discussed by us earlier.~\cite{Burkov08}
We expect the same to be true in the case of the incommensurate CDW dynamical instability.

\section{Experimental Implications}

Assuming that future experiments will be able to produce attractive cold Fermi gases in the lowest band of an 
optical lattice, it is reasonable to expect that the ground state and low energy excitations of such a system
will be well described by the attractive Hubbard model.\cite{lehur09,levin}
Our results on the collective mode excitations of the Hubbard model, specifically the roton gap and the
sound velocity as a function of filling and interaction strength, could then be verified experimentally.

Turning to superflow instabilities,  the theoretically predicted Landau and dynamical instabilities of Bose
superfluids \cite{Niu01,Altman05,nikuni06}
have been experimentally observed using cold bosonic atoms in an optical lattice.~\cite{ketterle-bose07}
As shown there, a flow in a cold atom system can be 
induced by frequency-detuning one pair of the counterpropagating laser beams that form the lattice. This creates a ``running" optical lattice 
potential in the corresponding basis direction, i.e. the lattice is effectively moving with respect to the stationary potential of the trap, holding 
the atomic cloud.
The effect of this on the bosons is easy to understand if one goes to the reference frame of the moving optical lattice. 
In this frame, the bosons must condense in a state of non-zero lattice momentum. If the optical lattice is 
moved sufficiently slowly, we expect the initial zero-momentum SF ground state of the bosons to adiabatically 
acquire a phase gradient, thus increasing the superflow momentum until dissipation sets in and destroys the 
condensate upon crossing a Landau or dynamical instability.

The instabilities of superflow for fermions loaded in an optical lattice may similarly be tested in cold atom 
experiments. This has been clearly demonstrated
in a pioneering experiment of Ref.~[\onlinecite{Ketterle07}] which studied the superfluid phase of fermions
in a harmonic trap and showed that the critical velocity smoothly interpolates between the pair breaking
velocity in the BCS regime to the Landau critical velocity in the BEC regime.~\cite{sensarma08} In a deep optical lattice, provided
that the superflow does not lead to excitations into the higher bands,
the superfluid phase
should be reasonably well described by the Hubbard model in the lowest 
band. In
this case, one may be able to probe the various dynamical phase diagrams proposed in this paper. Even
before reaching the instability, one may be able to use Bragg spectroscopy \cite{aspect03} to probe the 
collective density fluctuation spectrum, and observe the roton softening with increasing lattice velocity 
which would enable one to test the mechanism by which the superflow breaks down.

Our results are also of relevance for the vortex-core physics in strongly-correlated lattice superfluids. 
Considering the vortex core as simply the region where the current has exceeded the  value beyond which 
the uniform superfluid phase is no longer stable, we would expect the dynamical instabilities discussed in 
this paper to lead to commensurate or incommensurate density orders in the vortex core.

\section{Conclusions}
In conclusion, we have studied the SF ground state of the negative-U Hubbard model on a square and cubic lattice.
Using GRPA, we have studied the collective mode spectrum at all values of the interaction strength and density. 
We have shown that strong density correlations at increasing filling lead to the appearance of roton features in the 
collective mode spectrum, which signal developing competition between uniform SF and density-ordered CDW states. 
When superflow is imposed on the system, depairing and Landau instabilities arise. In addition, this competition 
leads to dynamical instabilities corresponding to 
roton mode softening. This dynamical instability occurs at a commensurate wavevector 
corresponding to the corner of the Brillouin zone. This instability is easily described in the strong coupling pseudospin 
language and is a consequence of the degeneracy between SF and 
``checkerboard" CDW ground states at $f=1$. 
Another very interesting dynamical instability occurs at an incommensurate wavevector and has no strong coupling analog. 
It occurs over a range of densities near $f=1/2$ and at intermediate values of the interaction strength. 
Our results may be verified in experiments with ultracold fermions loaded on an optical lattice.
Interesting avenues to explore in the future would include extending this work to other lattice
geometries~\cite{zhao06} and to study superflow instabilities in multiband superfluids.~\cite{burkov09}

\acknowledgments
This research was supported by NSERC of Canada. AP acknowledges support from the Sloan Foundation,
the Ontario ERA, and the Connaught Foundation. We thank the authors of Ref.[\onlinecite{danshita}] for 
pointing out an error in Fig.~7 in a previous version.

\appendix

\section{Fourier transform conventions}
We have Fourier transformed fermion operators as
\begin{eqnarray}
 c_{i\sigma} = \frac{1}{\sqrt{N}}\sum_{\bf{k}} c_{\bf{k} \sigma}e^{i\bf{k}.\bf{r}_i}.
\end{eqnarray} 
This defines the Fourier transform relations of our modulation operators as 
\begin{eqnarray}
\hat{\rho}_{i} = \frac{1}{N}\sum_{\bf{K}} \hat{\rho}_{\bf{K}} e^{i\bf{K}.\bf{r}_{i}}, \\
\hat{\Delta}_{i} = \frac{1}{N}\sum_{\bf{K}} \hat{\Delta}_{\bf{K}} e^{i\bf{K}.\bf{r}_{i}}, \\
\end{eqnarray}
where the momentum space operators $\hat{\rho}_{\bf{K}}$ and $\hat{\Delta}_{\bf{K}}$ are given in Eq.(\ref{eq:17}).

Also, the real space fields that couple to our modulation operators are Fourier transformed as

\begin{eqnarray}
h_{\rho/\Delta} (i,t) = \frac{1}{N} \int \frac{d\omega}{2\pi} \sum_{\bq} h_{\rho/\Delta}(\bq,\omega) 
e^{i(\bq\cdot\br_i-\omega t)}.
\end{eqnarray}

These Fourier transforms have been used in going from Eq.(\ref{eq:14}) to Eq.(\ref{eq:16}).

\section{Details of the Generalized Random Phase Approximation}
Let us assume that we have a Hamiltonian $H_0$ which is
modified by a set of weak time-dependent perturbations
so that
\bea
\label{hpert}
H(t)=H_0 - \frac{1}{N}
\sum_{\bK} h_\alpha(\bK,t) \hat{O}^\dg_{\alpha}(\bK),
\eea
where the perturbation is assumed to be Hermitian.
Using time-dependent perturbation theory to leading order in the perturbing 
fields, one finds that the change in the expectation value of the operator $O_{\alpha}(\bK)$ is 
given by:
\be
\delta \la \hat{O}_\alpha \ra (\bK,t) = \int_{-\infty}^{+\infty} dt'~\chi^0_{\alpha\beta}(\bK,t - t') h^\pdg_\beta(\bK,t'),
\ee
where
\be
\chi^0_{\alpha\beta}(\bK,t\!-\!t') \!=\! i \Theta(t\!-\!t') \la 
\left[ \hat{O}^\pdg_\alpha(\bK,t), \hat{O}^\dg_\beta(\bK,t')\right] \ra_{_0}.
\ee
Here $\left[ . , .\right]$ denotes the commutator and $\la . \ra_0$ implies
that the expectation value is taken in
the ground state of $H_0$.
Upon doing a spectral decomposition, one obtains:
\be
\chi^0_{\alpha\beta}(\bK,\omega)
\!=\!\!\frac{1}{N}\!\!\sum_n \!\!\left(\! \frac{(\hat{O}^\dg_\beta)_{0n}
(\hat{O}^\pdg_\alpha)_{n0}}{\omega\!\!+\!\!E_{n0}\!\!+\!\!i 0^+}
\!-\! \frac{(\hat{O}^\pdg_\alpha)_{0n}
(\hat{O}^\dg_\beta)_{n0}}{\omega\!\!-\!\!E_{n0}\!\!+\!\!i 0^+}\!\!\right).
\ee
Here $(\hat{O})_{mn} \equiv \la m | \hat{O} | n \ra$, and $| n \ra, |m\ra$
denote the eigenstates of $H_0$ (with $n\!\!=\!\!0$ corresponding to the ground state).
In the denominator, $E_{n0} \equiv E_n\!\!-\!\!E_0$ where
$E_n$ is the energy of state $|n\ra$.

We are interested in the case where $H_0=H_{\rm MFT}$, 
with the operators 
$\hat{O}^\dg_\alpha(\bK)$ given by
\begin{eqnarray}
\hat{O}^\dg_1 (\bK) &=& \hat{\rho}^\pdg_{-\bK}, \nonumber  \\
\hat{O}^\dg_2 (\bK) &=& \hat{\Delta}^\pdg_{-\bK+\bQ}, \nonumber  \\
\hat{O}^\dg_3 (\bK) &=& \hat{\Delta}_{\bK+\bQ}^\dagger,
\end{eqnarray}
and the perturbing fields corresponding to
\bea
h_1(\bK,t)&=&h_\rho(\bK,t), \nonumber  \\
h_2(\bK,t)&=&h_\Delta(\bK,t), \nonumber \\
h_3(\bK,t)&=&h_\Delta^*(-\bK,t).
\eea
At zero temperature, it is easily checked that the quasiparticle energies are all positive
in the dynamically stable region of interest. In this case,
the only terms which have nonzero matrix
elements in the expression for $\chi^0_{\alpha\beta}(\bK,\omega)$ are those where the operators
create two Bogoliubov quasiparticles (QPs) when acting on $|0\ra$ and where they destroy two
Bogoliubov QPs when acting on $| n\ra$.
It is therefore convenient to resolve each perturbation operator into two parts --
one that creates two QPs and one that annihilates two QPs -- as follows:
\bea
\hat{\rho}^c_{-\bK}\!\!&\!=\!&\!\!\frac{1}{2} \!\sum_\bk \!(u_{\bk\!+\!\bK} v_{\bk}\! +\! u_\bk v_{\bk\!+\!\bK}\!) 
\gamma^\dg_{\bk\!+\!\bK\upa} \gamma^\dg_{-\bk\!+\!\bQ\dna}, \nonumber  \\
\hat{\rho}^a_{-\bK} \!\!&\!=\!&\!\!\frac{1}{2} \!\sum_\bk \!(u_{\bk\!-\!\bK} v_{\bk}\! +\! u_\bk v_{\bk\!-\!\bK}\!) 
\gamma^\pdg_{-\bk\!+\!\bQ\dna} \gamma^\pdg_{\bk\!-\!\bK\upa}, \nonumber \\
\hat{\Delta}^c_{-\bK+\bQ}\!\!&\!=\!&\!\! -\!\sum_\bk\! v_\bk v_{\bk+\bK} \gamma^\dg_{\bk+\bK\upa}
\gamma^\dg_{-\bk+\bQ\dna}, \nonumber \\
\hat{\Delta}^a_{-\bK+\bQ}\!\!&\!=\!&\!\! \sum_\bk\! u_\bk u_{\bk - \bK} \gamma^\pdg_{-\bk+\bQ\dna}
\gamma^\pdg_{\bk-\bK\upa}, \nonumber \\
\hat{\Delta}^{\dg c}_{\bK+\bQ}\!\!&\!=\!&\!\! \sum_\bk\! u_\bk u_{\bk+\bK} \gamma^\dg_{\bk+\bK\upa}
\gamma^\dg_{-\bk+\bQ\dna}, \nonumber \\
\hat{\Delta}^{\dg a}_{\bK+\bQ}\!\!&\!=\!&\!\! \!-\!\sum_\bk\! v_\bk v_{\bk-\bK} \gamma^\pdg_{-\bk+\bQ\dna}
\gamma^\pdg_{\bk-\bK\upa}.
\eea
where the superscript `c' denotes creation of 2 quasiparticles, and `a' denotes annihilation.

Since $\chi^0$ is a symmetric matrix, it suffices to use the above to compute the following
distinct elements:
\bea
\chi^{0}_{1,1}=\frac{1}{4N} \sum_{\bk} &&\!\!\!\!\!\!\!\!\!\!\!\! 
\left[
\frac{(u_{\bk-\bK}v_{\bk}+v_{\bk-\bK}u_{\bk})^2} {\omega+E_{\bk-\bK}+E_{-\bk+\bQ}} \right.\nn\\
&-& \left. \frac{(u_{\bk}v_{\bk+\bK}+v_{\bk}u_{\bk+\bK})^2}{\omega-E_{\bk+\bK}-E_{-\bk+\bQ}}\right], 
\nonumber \\
\chi^{0}_{1,2}=\frac{1}{2N} \sum_{\bk} && \!\!\!\!\!\!\!\!\!\!\!\! 
\left[\frac{(u_{\bk-\bK}v_{\bk}+v_{\bk-\bK}u_{\bk}) u_\bk u_{\bk-\bK}} {\omega+E_{\bk-\bK}+E_{-\bk+\bQ}} \right.\nn\\
\!\!\!\!\!\!\!\!\!\!\!\!&+&\!\!\!\!\left. \frac{(u_{k}v_{k+K}\!\!+\!\!v_{k}u_{k+K}) v_k v_{k+K}}
{\!\omega\!-\!E_{\bk+\bK}\!-\!E_{-\bk+\bQ}}\!\right], \nonumber  \\
\chi^{0}_{1,3}=\frac{1}{2N} \sum_{\bk} && \!\!\!\!\!\!\!\!\!\!\!\! 
\left[-\frac{(u_{\bk-\bK}v_{\bk}+v_{\bk-\bK}u_{\bk}) v_\bk v_{\bk-\bK}}{\omega+E_{\bk-\bK}+E_{-\bk+\bQ}} \right.\nn\\
\!\!\!\!\!\!\!\!\!\!\!\!\!\!\!&-& \!\!\!\!\left. \frac{(u_{\bk}v_{\bk+\bK}\!\!+\!\!v_{\bk}u_{\bk+\bK}) u_\bk u_{\bk+\bK}}
{\omega\!\!-\!\!E_{\bk+\bK}\!\!-\!\!E_{-\bk+\bQ}}\!\right], \nonumber \\
\chi^{0}_{2,2}= ~\frac{1}{N} \sum_{\bk} && \!\!\!\!\!\!\!\!\!\!\!\! 
\left[\frac{u^2_{\bk} u^2_{\bk-\bK}}{\omega+E_{\bk-\bK}+E_{-\bk+\bQ}} \right.\nn\\
&-&\left. \frac{v^2_{\bk}v^2_{\bk+\bK}}
{\omega-E_{\bk+\bK}-E_{-\bk+\bQ}}\right], \nonumber \\
\chi^{0}_{2,3}=\frac{1}{N} \sum_{\bk} &&\!\!\!\!\!\!\!\!\!\!\!\! 
\left[- \frac{u_{\bk} v_\bk u_{\bk-\bK} v_{\bk-\bK}}
{\omega+E_{\bk-\bK}+E_{-\bk+\bQ}} \right.\nn\\
&+& \left. \frac{u_\bk v_{\bk} u_{\bk+\bK} v_{\bk+\bK}}
{\omega-E_{\bk+\bK}-E_{-\bk+\bQ}}\right], \nonumber \\
\chi^{0}_{3,3}=\frac{1}{N} \sum_{\bk} && \!\!\!\!\!\!\!\!\!\!\!\!
\left[\frac{v^2_{\bk} v^2_{\bk-\bK}}{\omega+E_{\bk-\bK}+E_{-\bk+\bQ}} \right.\nn\\
&-& \left. \frac{u^2_{\bk}u^2_{\bk+\bK}}
{\omega-E_{\bk+\bK}-E_{-\bk+\bQ}}\right]. 
\eea

In order to include interaction effects within the GRPA, we note that the interaction
can be decomposed as follows:
\begin{eqnarray}
\!\!\!\!\!\!\!- U c_{i\uparrow}^\dagger c_{i\downarrow}^\dagger c_{i\downarrow} c_{i\uparrow}
\!\!\rightarrow\!\!\!&-& \!\!\!U\!\!\left[\!
\langle c_{i\uparrow}^\dagger c_{i\uparrow}\rangle c_{i\downarrow}^\dagger c_{i\downarrow} \!\!+ \!\!\langle c_{i\downarrow}^\dagger c_{i\downarrow}\rangle c_{i\uparrow}^\dagger c_{i\uparrow}\!\right] \nn \\
&-&\!\!\!U\!\left[\! 
\langle c_{i\uparrow}^\dagger c_{i\downarrow}^\dagger\rangle c_{i\downarrow}c_{i\uparrow}\!\!+\!\!\langle c_{i\downarrow} c_{i\uparrow}\rangle c_{i\uparrow}^\dagger c_{i\downarrow}^\dagger \!\right]
\end{eqnarray}
These expectation values, generated by interactions in the presence of  external fields, act
as ``internal fields"  which renormalize the applied field. It is easy to show that this effect
can be taken into account by simply setting 
\bea
h_1(\bK,\omega) &\to& h_1(\bK,\omega) + 2 U \la\hat{O}_1(\bK,\omega)\ra, \nonumber \\
h_2(\bK,\omega) &\to& h_2(\bK,\omega) + U \la\hat{O}_2(\bK,\omega)\ra, \nonumber \\
h_3(\bK,\omega) &\to& h_3(\bK,\omega) + U \la\hat{O}_3(\bK,\omega)\ra.
\eea
This leads to:
\be
\delta \la\hat{O}_\alpha\ra = \chi^0_{\alpha\beta} (h_\beta +
U D_{\beta\tau} \delta \la\hat{O}_\tau\ra),
\ee
where $D\equiv {\rm Diag}\{2,1,1\}$  is a diagonal matrix, and we have suppressed
$(\bK,\omega)$ labels for notational simplicity.
Solving the above equation gives:
\begin{eqnarray}
\delta \langle \hat{O}_\alpha \rangle = [(1-U \chi^{0} D )^{-1}\chi^{0}]_{\alpha \beta}~h_{\beta} \equiv\chi^{GRPA}_{\alpha \beta}h_{\beta},
\end{eqnarray}
where $\chi_{\alpha\beta}^{GRPA}$ is the renormalized GRPA susceptibility.

\section{Mean field theory of the flowing supersolid}

For all values of the interaction strength and for fermion densities near $f=1$, we clearly see a dynamical instability at the commensurate ``checkerboard" 
wavevector - $(\pi,\pi)$ in 2D and $(\pi,\pi,\pi)$ in 3D. At this instability, we expect ``checkerboard" density order to arise beyond critical flow,
leading to 
a state with coexisting SF and density order, a ``flowing supersolid".~\cite{zhao05,Burkov08}  We study the
mean field theory of this state in order to examine its stability.
In the presence of a nonzero
SF flow, imposed as a uniform phase gradient on $\Delta$, 
we self-consistently calculate the ground state allowing for ``checkerboard" 
density modulations. 

The mean field order parameters are: 
\begin{eqnarray}
\Delta &\equiv& \frac{U}{N} \sum_{\bk} \la c_{-\bk+\bQ\downarrow}c_{\bk\uparrow} \ra, \nonumber  \\
\tilde{\rho} &\equiv& \frac{U}{2 N}  \sum_{\bk} \la c_{\bk+\bPi\sigma}^\dagger c^\pdg_{\bk\sigma} \ra, \nonumber  \\
\tilde{\Delta}&\equiv& \frac{U}{N} \sum_{\bk } \la c_{-\bk+\bPi+\bQ\downarrow}c_{\bk\uparrow}\ra,
\end{eqnarray}
where $\bPi \equiv (\pi,\pi)$ in 2D or $(\pi,\pi,\pi)$ in 3D. 
Since there are no spin-selective Zeeman terms, we are justified in forcing the density modulation to be equal for both spin species. 
Due to global phase rotation U(1) symmetry, we can choose
$\Delta$ to be real but we allow $\tilde{\Delta}$ to be complex. 
$\tilde{\rho}$, being the expectation value of the staggered density, is real.

Upto an unimportant constant, the Hamiltonian can then be written as: 
\begin{equation}
H = \sum_{\bk}{}^{'}\Psi^\dagger_\bk \underline{H}(\bk) \Psi^\pdg_\bk,
\end{equation}
where the primed summation in the Hamiltonian indicates that if $\bk$ is included, then $\bk+\bPi$ is to be excluded. 
The other notation is as follows:
\begin{eqnarray}
\Psi_{\bk} = \left(\begin{array}{c}
       c_{\bk\uparrow} \\
	c_{-\bk+\bQ\downarrow}^\dagger \\
	c_{\bk+\bPi\uparrow} \\
	c_{-\bk-\bPi+Q\downarrow}^\dagger
      \end{array}\right),
\end{eqnarray}
and
\begin{eqnarray}
\label{4x4hmlt}
\underline{H} =\left(\begin{array}{cccc}
       \xi_\bk &	-\Delta & -\tilde{\rho} & -\tilde{\Delta} \\
	-\Delta & -\xi_{-\bk+\bQ} &	-\tilde{\Delta}^* & \tilde{\rho} \\
	-\tilde{\rho} &	-\tilde{\Delta} & \xi_{\bk+\bPi} &	-\Delta \\
	-\tilde{\Delta}^{*} & \tilde{\rho} & -\Delta & -\xi_{-\bk-\bPi+\bQ}
      \end{array}\right). 
\end{eqnarray}

For given $U$, $\bQ$ and the density, we numerically diagonalize this matrix and solve the self-consistency equations 
for $\Delta$, $\tilde{\rho}$, $\tilde{\Delta}$ and the filling $f$. 
We then evaluate the superfluid and density order parameters in the converged solution.
In addition, we also evaluate the uniform current,
\begin{equation}
\langle\hat{\cal J}\rangle=-2t\langle \sum_{\bk} c_{\bk\sigma}^\dagger c^\pdg_{\bk\sigma} {\boldsymbol \nabla}_{\bk} \epsilon_{\bk} \rangle
\end{equation}
where $-2t\epsilon_\bk$ is the non-interacting fermion dispersion.

\begin{figure}
\includegraphics[width=2.8in]{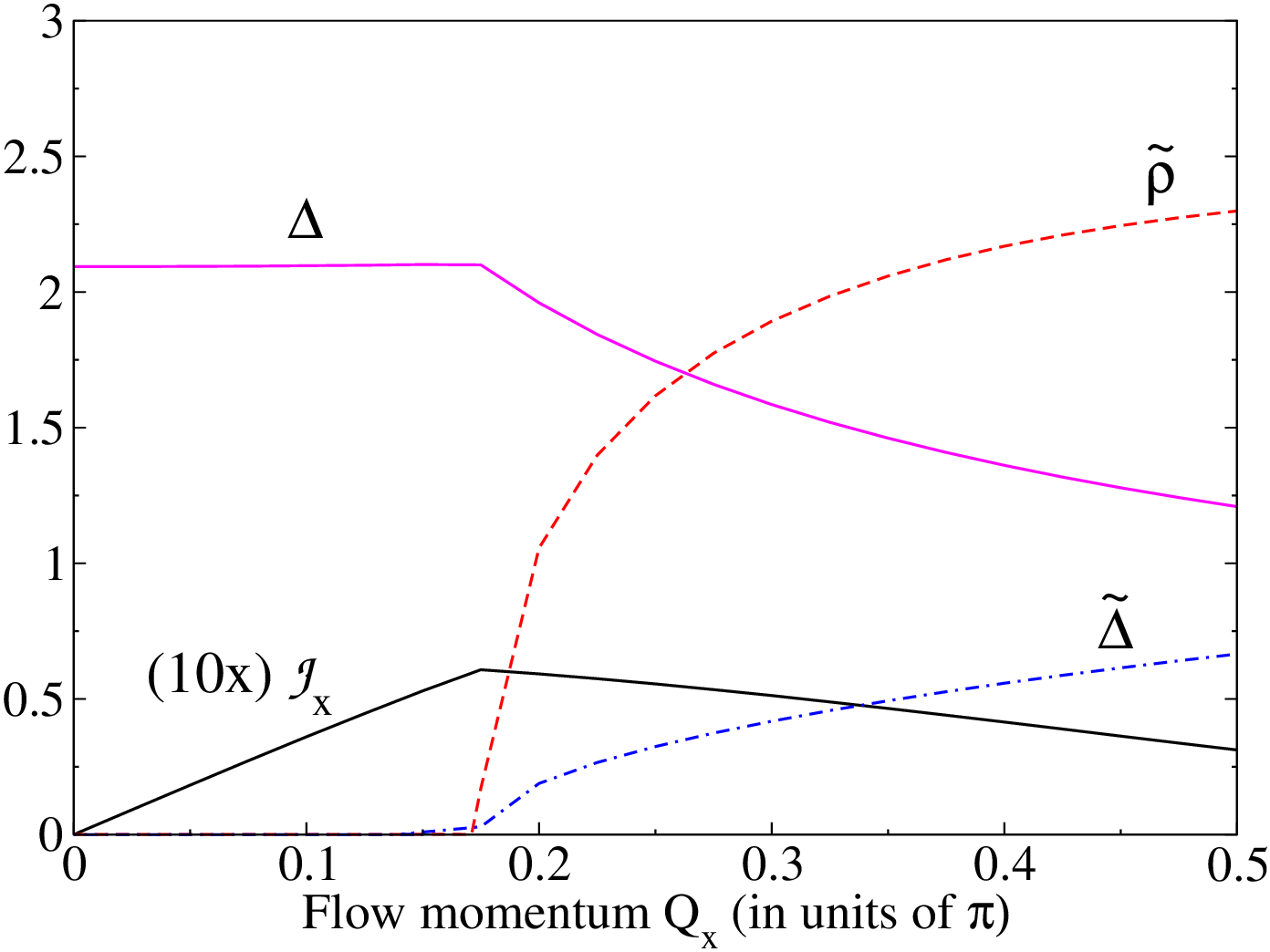}
\caption{(Color online)
Mean field theory results for the ``flowing supersolid" state showing the
various observables as functions of the flow momentum in 2D for U/t=7 and with a filling of 0.8 fermions 
per site. Supersolid order onsets around $Q_x \approx 0.2 \pi$, as both $\Delta$ and $\tilde{\rho}$ have simultaneous non-zero expectation values. This coincides with a maximum in the current as a function of the flow momentum, indicating a dynamical instability.} 
\label{sqmftpipi}
\end{figure}

A ``flowing supersolid" phase is indicated by simultaneous non-zero values for $\Delta$ and $\tilde{\rho}$. For the case of checkerboard order
we do see a ``flowing supersolid" phase beyond a critical flow momentum.
However, the onset of density order coincides with a maximum in the expectation value of the current as a function of flow momentum(Fig.~\ref{sqmftpipi}). This indicates that the system is dynamically unstable to the exponential
growth phase and density fluctuations. The argument below demonstrates why this is the case in a simpler
context.

Let us consider a superfluid system in 1D for simplicity. Denoting the mean field density and phase of the SF order 
parameter by $n_0$ and $\phi_0$ respectively, we consider fluctuations $\delta n$ and $\delta \phi$. Now, the current is some function of the 
gradient of the phase: 
\begin{equation}
 \la \hat{\cal J} \ra = {\cal J}\left(\frac{d\phi}{dx}\right).
\end{equation}
The equations governing the dynamics of the fluctuations are the Josephson relation and the continuity equation. The former gives:
\begin{eqnarray}
\frac{d\delta\phi}{dt}=-\alpha \delta n, 
\end{eqnarray}
where $\alpha = d\mu/dn > 0$ is closely related to the compressibility. The continuity equation is:
\begin{equation}
\nonumber \frac{d n}{dt}=-\frac{d{\cal J}}{dx}.
\end{equation}
Substituting $n = n_0+\delta n$ and $\frac{d \phi}{d x}  = Q + \frac{d \delta \phi}{d x}$, where $Q \equiv \frac{d \phi_0}{d x}$, we obtain 
\begin{equation}
\frac{d\delta n}{dt}= -\frac{d{\cal J}}{dQ} \frac{d^2 \delta\phi}{dx^2}.
\end{equation}
Combining this with the Josephson relation, we finally obtain:
\begin{equation}
\frac{1}{\alpha}\frac{d^2\delta\phi}{dt^2}=\frac{d{\cal J}}{dQ}\frac{d^2 \delta \phi}{dx^2}
\end{equation}
When $d{\cal J}/dQ$, becomes negative, the wavelike solutions of this equation develop complex frequencies. In this case, density and phase fluctuations will grow exponentially in time making the system dynamically unstable 
when the current goes through a maximum as a function of the flow momentum Q.


\begin{thebibliography}{99}
\bibitem{WenLee} P.A. Lee, N. Nagaosa, and X.-G. Wen, Rev. Mod. Phys. {\bf 78}, 17 (2006).
\bibitem{norman08}
M. R. Norman, Physics {\bf 1}, 21 (2008).
\bibitem{Bloch08} I. Bloch, J. Dalibard, and W. Zwerger, Rev. Mod. Phys. {\bf 80}, 885 (2007). 
\bibitem{lehur09}
Karyn Le Hur and T. Maurice Rice, Annals of Physics {\bf 324}, 1452 (2009).
\bibitem{Aeppli01} B. Lake, G. Aeppli, K.N. Clausen, D.F. McMorrow, K. Lefmann, N.E. Hussey, N. Mangkorntong, M. Nohara, 
H. Takagi, T.E. Mason, and A. Schroder, Science {\bf 291}, 1759 (2001). 
\bibitem{Niu01} B. Wu and Q. Niu, Phys. Rev. A {\bf 64}, 061603 (2001). 
\bibitem{Altman05} E. Altman,  A. Polkovnikov, E. Demler, B.I. Halperin, and M.D. Lukin, Phys. Rev. Lett. {\bf 95}, 020402 (2005); A. Polkovnikov, E. Altman, E. Demler, B. Halperin, M.D. Lukin, \pra {\bf 71}, 063613 (2005).
\bibitem{nikuni06}
K. Iigaya, S. Konabe, I. Danshita, and T. Nikuni, \pra {\bf 74}, 053611 (2006).
\bibitem{ketterle-bose07}
J. Mun, P. Medley, G. K. Campbell, L. G. Marcassa, D. E. Pritchard, W. Ketterle,
\prl {\bf 99}, 150604 (2007).
\bibitem{sensarma06} R. Sensarma, M. Randeria, and T.-L. Ho, \prl {\bf 96},
090403 (2006).
\bibitem{Ketterle07} D.E. Miller, J.K. Chin, C.A. Stan, Y. Liu, W. Setiawan, C. Sanner, and W. Ketterle, Phys. Rev. Lett. {\bf 99}, 070402 (2007).
\bibitem{Ranninger90} R. Micnas, J. Ranninger, and S. Robaszkiewicz, Rev. Mod. Phys. {\bf 62}, 113 (1990). 
\bibitem{levin}
C.-C. Chien, Y. He, Q. Chen, and K. Levin, \pra {\bf 77}, 011601 (2008); C.-C. Chen, Q. Chen,
and K. Levin, \pra {\bf 78}, 043612 (2008).
\bibitem{swz}
D. J. Scalapino, S. R. White, and S. C. Zhang, \prl {\bf 68}, 2830 (1992); D. J. Scalapino,
S. R. White, and S. C. Zhang, \prb {\bf 47}, 7995 (1993).
\bibitem{Scalapino91} A. Moreo and D.J. Scalapino, Phys. Rev. Lett. {\bf 66}, 946 (1991). 
\bibitem{trivedi}
M. Randeria, N. Trivedi, A. Moreo, and R. T. Scalettar, \prl {\bf 69}, 2001 (1992);
N. Trivedi and M. Randeria, \prl {\bf 75}, 312 (1995).
\bibitem{Zhang90} S.-C. Zhang, Phys. Rev. Lett. {\bf 65}, 120 (1990). 
\bibitem{MacDonald08} W.-C. Lee, J. Sinova, A.A. Burkov, Y. Joglekar, and A.H. MacDonald, Phys. Rev. B {\bf 77}, 214518 (2008). 
\bibitem{morosan06} E. Morosan, H. W. Zandbergen, B. S. Dennis, W. G. Bos, Y. Onose,
T. Klimczuk, A. P. Ramirez, N. P. Ong, R. J. Cava, Nature Physics {\bf 2} 544 (2006).
\bibitem{kostyrko}
T. Kostyrko and R. Micnas, Phys, Rev. B {\bf 46}, 11025 (1992).
\bibitem{alm}
T. Alm
and P, Schuck, Phys. Rev. B {\bf 54}, 2471 (1996).
\bibitem{danshita}
Y. Yunomae, I. Danshita, D. Yamamoto, N. Yokoshi, S. Tsuchiya, arXiv:0809.0350 (unpublished);
Y. Yunomae, D. Yamamoto, I. Danshita, N. Yokoshi, S. Tsuchiya, arXiv:0904.3179 (unpublished).
\bibitem{Burkov08} A.A. Burkov and A. Paramekanti, Phys. Rev. Lett. {\bf 100}, 255301 (2008). 
\bibitem{Halperin1968} B.I. Halperin,, and T.M. Rice, Rev. Mod. Phys. {\bf 40}, 755 (1968).
\bibitem{Anderson58} P.W. Anderson, Phys. Rev. {\bf 112}, 1900 (1958). 
\bibitem{auerbach}
{\em Interacting Electrons and Quantum Magnetism}, A. Auerbach (Springer-Verlag, New York, 1994).
\bibitem{Randeria94} L. Belkhir and M. Randeria, Phys. Rev. B {\bf 49}, 6829 (1994). 
\bibitem{volovik}
{\em Universe in a Helium Droplet}, G. E. Volovik (Oxford University Press, Oxford, 2003).
\bibitem{Wei09} See T.-C. Wei and P.M. Goldbart, arXiv:0904.2409 and references therein.  
\bibitem{sensarma08}
R. B. Diener, R. Sensarma, and M. Randeria,
\pra {\bf 77}, 023626 (2008).
\bibitem{aspect03}
S. Richard, F. Gerbier, J. H. Thywissen, M. Hugbart, P. Bouyer, A. Aspect
\prl {\bf 91}, 010405 (2003).
\bibitem{zhao06}
E. Zhao and A. Paramekanti, \prl {\bf 97}, 230404 (2006).
\bibitem{burkov09}
A. A. Burkov and A. Paramekanti, \pra {\bf 79}, 043626 (2009).
\bibitem{zhao05} E. Zhao and A. Paramekanti,
\prl {\bf 96}, 105303 (2006).
\end{thebibliography}
\end{document}